\documentclass[aps,twocolumn,superscriptaddress]{revtex4-1}
\usepackage{graphicx}
\usepackage{bm}
\usepackage{hyperref}
\usepackage{color}
\usepackage{amsfonts}
\usepackage{amsmath}
\usepackage{tikz-cd}
\usepackage{tikz}
\usepackage{pgfplots}
\usepackage{epstopdf}

\newcommand{\ocaaddress}{Laboratoire Lagrange, Universit\'e C\^ote
  d'Azur, CNRS, OCA, Bd.\ de l'Observatoire, Nice, France}
\newcommand{\cemefaddress}{MINES ParisTech, Center for Materials
  Forming (CEMEF), CNRS UMR\,7635, Sophia Antipolis, France }

\begin{document}

\title{Inelastic accretion of inertial particles by a towed sphere}

\author{Robin Vall\'ee}
\affiliation{\cemefaddress}
\affiliation{\ocaaddress}
\author{Christophe Henry}
\affiliation{\ocaaddress}
\author{Elie Hachem}
\affiliation{\cemefaddress}
\author{J\'er\'emie Bec}
\affiliation{\ocaaddress}

\begin{abstract}
  The problem of accretion of small particles by a sphere embedded in
  a mean flow is studied in the case where the particles undergo
  inelastic collisions with the solid object. The collision
  efficiency, which gives the flux of particles experiencing at least
  one bounce on the sphere, is found to depend upon the sphere
  Reynolds number only through the value of the critical Stokes number
  below which no collision occurs. In the absence of molecular
  diffusion, it is demonstrated that multiple bounces do not provide
  enough energy dissipation for the particles to stick to the surface
  within a finite time.  This excludes the possibility of any kind of
  inelastic collapse, so that determining an accretion efficiency
  requires modelling more precisely particle-surface microphysical
  interactions. A straightforward choice is to assume that the
  particles stick when their kinetic energy at impact is below a
  threshold.  In this view, numerical simulations are performed in
  order to describe the statistics of impact velocities at various
  values of the Reynolds number. Successive bounces are shown to
  enhance accretion. These results are put together in order to
  provide a general qualitative picture on how the accretion
  efficiency depends upon the non dimensional parameters of the
  problem.
\end{abstract}

\maketitle

\section{Introduction}

A number of situations involve interactions between small particles
suspended in a fluid and a boundary. These include industrial
applications, such as sediment deposition in ducts and
fouling~\cite{henry2012towards}, but also natural phenomena such as
aerosol scavenging by raindrops~\cite{pruppacher-klett:1997} or
accretion of dust onto planetary embryos~\cite{lissauer:1993}.  Recent
studies have shown that the combined effects of particles inertia and
of inelastic shocks among them or with a surface leads to intricate
outcomes.  For instance, particles can undergo sticky elastic
collisions and form clusters under the sole influence of their
dissipative viscous drag~\cite{bec2013sticky}. Moreover, it was shown
that there exists a localization/delocalization transition depending
on their Stokes number and their restitution coefficient that rules
the long-time inhomogeneities in the particles spatial
distribution~\cite{belan2014localization}. This approach, which was
initially focusing on random fluid flows vanishing linearly at the
boundary, has recently been extended to nonlinear flows mimicking the
behavior in viscous boundary
layers~\cite{belan2016concentration,belan2016inelastic}.  The
mechanisms at play are very similar to those ruling the inelastic
collapse of randomly forced particles~\cite{cornell1998inelastic}, but
are in that case extended to space-dependent diffusion coefficients.
Such a similarity raises two questions. First, one can wonder whether
or not a noisy behavior in the particles dynamics is absolutely
necessary for observing a collapse to the surface. Thermal-like
agitation could indeed be key in activating such a transition, as
observed in granular dynamics~\cite{cecconi2003noise}.  Conversely a
deterministic behavior of the fluid flow close to the boundary,
inducing for instance a constant drift that pushes the particles
toward the surface, could be an effective way to trigger collapse. A
second observation is that inelastic collapse typically occurs in a
finite time, so one can conjecture that particles-wall interactions
affect not only the stationary long-time distribution but also
transient dynamics. This could have noticeable consequences on the
question of particle accretion and deposition on a spherical obstacle.

Rain is known to give an important contribution to the transfer of
aerosol particles onto the ground. This ``wet deposition'' originates
from the scavenging of particles by drops during their precipitation
and is a key ingredient entering cloud-resolving meteorological and
climatic models~\cite{mircea2000precipitation,berthet2010scavenging}.
Aerosols play a central role in atmospheric physics, not only as cloud
condensation nuclei but also as hazardous particulate
pollutants. Accurate predictions on their lifecycle are thus needed to
efficiently estimate both long-term global
warming~\cite{levy2013roles} and pollution washout depending on
meteorological conditions~\cite{guo2016washout}. The modeling and
parametrization of scavenging is essentially built on the seminal work
by Beard~\cite{beard1974experimental} and on the numerical
investigation of the collection efficiency of small inertial particles
by a sphere in axisymmetric flows~\cite{beard1974numerical}.  Such
models are still subject to experimental
validations~\cite{ardon2015laboratory,lemaitre2017experimental}.
Accuracy is difficultly achievable because of the variety of physical
effects at play including diffusion, electro-scavenging due to
particle charges, thermophoresis stemming from the local air cooling
by the raindrop, and inertial impaction where sufficiently large
aerosols detach from the fluid streamlines to collide with the
drop. All these effects fail at providing a satisfactory mechanism for
scavenging aerosols with sizes $\simeq 0.1$ to $1\,\mu{\rm m}$.  Such
ultrafine particles are the most dangerous to health since they are
likely to penetrate the respiratory system up to the alveolar region
to provoke pulmonary, cardiovascular or brain
diseases~\cite{chen2016beyond}.  Under typical atmospheric conditions,
the collection efficiency is dominated by thermal diffusion at small
sizes and inertial impaction at larger and attains a minimum well
below 1\% at intermediate values corresponding to a particle Stokes
number (non-dimensional response time) of the order of $0.01$ to $1$.

The presence of a minimum of collection efficiency is above all due to
the ineffectiveness of inertial impaction at small particle sizes.
Particles with a too weak inertia tend to closely follow the fluid
velocity streamlines and are thus swept around the obstacle without
impacting it.  This effect acts drastically at small values of the
Stokes number: There actually exists a critical value below which no
particles collide with the sphere~\cite{phillips1999influence}.  Besides
leading to inefficiency in the accretion process, the existence of
this cutoff implies that the asymptotics of small Stokes numbers
becomes irrelevant, so that analytical results on inertial impaction
are particularly arduous.  This leaves the sole possibility of
employing empirical fitting formula, as for instance proposed by
Slinn~\cite{slinn1983precipitation}, in the models used to address
questions where inertial impaction is critical.  This is for instance
the case in astrophysical situations related to planetesimal growth by
dust accretion~\cite{sellentin2013quantification}. There, inertial
impaction is very important for the early stages of planetesimal
growth and might be triggered by the outer gas turbulence of the
protoplanetary disk~\cite{homann2016effect}. Most work has
nevertheless focused on the collision efficiency of dust with large
spherical bodies.  It is however known that in the context of dust
accretion, the outcome of a collision (coagulation, gravitational
capture, bounce, disruption, etc) depends on the details of the
impact, such as the kinetic energy, the collision angle and on the
material properties of the two bodies~\cite{chokshi1993dust}. Such
microphysical features need to be considered in population dynamics
models in order to properly predict the timescales of planet
growth~\cite{windmark2012breaking,garaud2013dust}. Previous work has
focused on the probability density of the normal component of impact
velocity at the first collision and obtained evidence of a Gaussian
distribution whose variance increases as a function of the Stokes
number~\cite{mitra2013can}. However, possible particle bouncing and
successive impacts have not yet been considered and could actually
lead to a significant depletion of collisional velocities and enhance
accretion.

Most previous work has focused on first-impact statistics. Here we
consider small inertial particles without molecular diffusion that
experience inelastic collisions with a large solid sphere maintained
fixed in a mean flow.  For that purpose, numerical simulations are
performed to investigate several issues related to the particle
accretion problem:
\begin{enumerate}
 \item[i.] How does the collision efficiency depend on well-chosen 
 dimensionless numbers (Stokes, Reynolds)?
 \item[ii.] Can particle bouncing and successive inelastic rebounds 
 enhance accretion (inelastic collapse)?
\end{enumerate}
We will see whether or not the actual effect of a deterministic flow
at the boundary is different from naive expectations that can be drawn
from existing works on inelastic collisions in simple idealized flows.

The paper is organized as follows: the accretion problem (collision
efficiency and outcome) is first discussed in
Section~\ref{sec:accretion_problem} together with a succint overview
of the model and simulation settings; we evaluate theoretically in
Section~\ref{sec:relax_succ_coll} whether successive bounces can lead
to inelastic collapse within a finite time; the statistics of impact
are analyzed numerically in Section~\ref{sec:impact_stats} to
characterize the role of successive impacts on accretion while
accounting for more realistic particle-sphere interactions.

\section{The accretion problem and model}
\label{sec:accretion_problem}

We consider the fluid flow past a fixed large sphere. This is a
representative case of all the axisymmetric bodies that are likely to
accrete small particles in the various situations described in the
previous section. The large sphere of diameter $d$ is centered at the
origin, and embedded in a flow prescribed to be $\bm u = U \bm e_z$ at
infinity and solving the incompressible Navier--Stokes equation with
no-slip boundary conditions at the sphere surface.  The flow is
characterized by the Reynolds number associated with the sphere
diameter $Re = U\,d/\nu$, where $\nu$ denotes the kinematic viscosity
of the fluid.  Numerical simulations are performed using a
second-order adaptive finite-element code~\cite{hachem2010stabilized}. 
Even if this code gives the
possibility to model turbulent flow using a variational multi-scale
method, we actually perform direct numerical simulations by
prescribing a sufficient resolution to accurately determine all
relevant scales.

\begin{figure}[h]
  \includegraphics[width=\columnwidth]{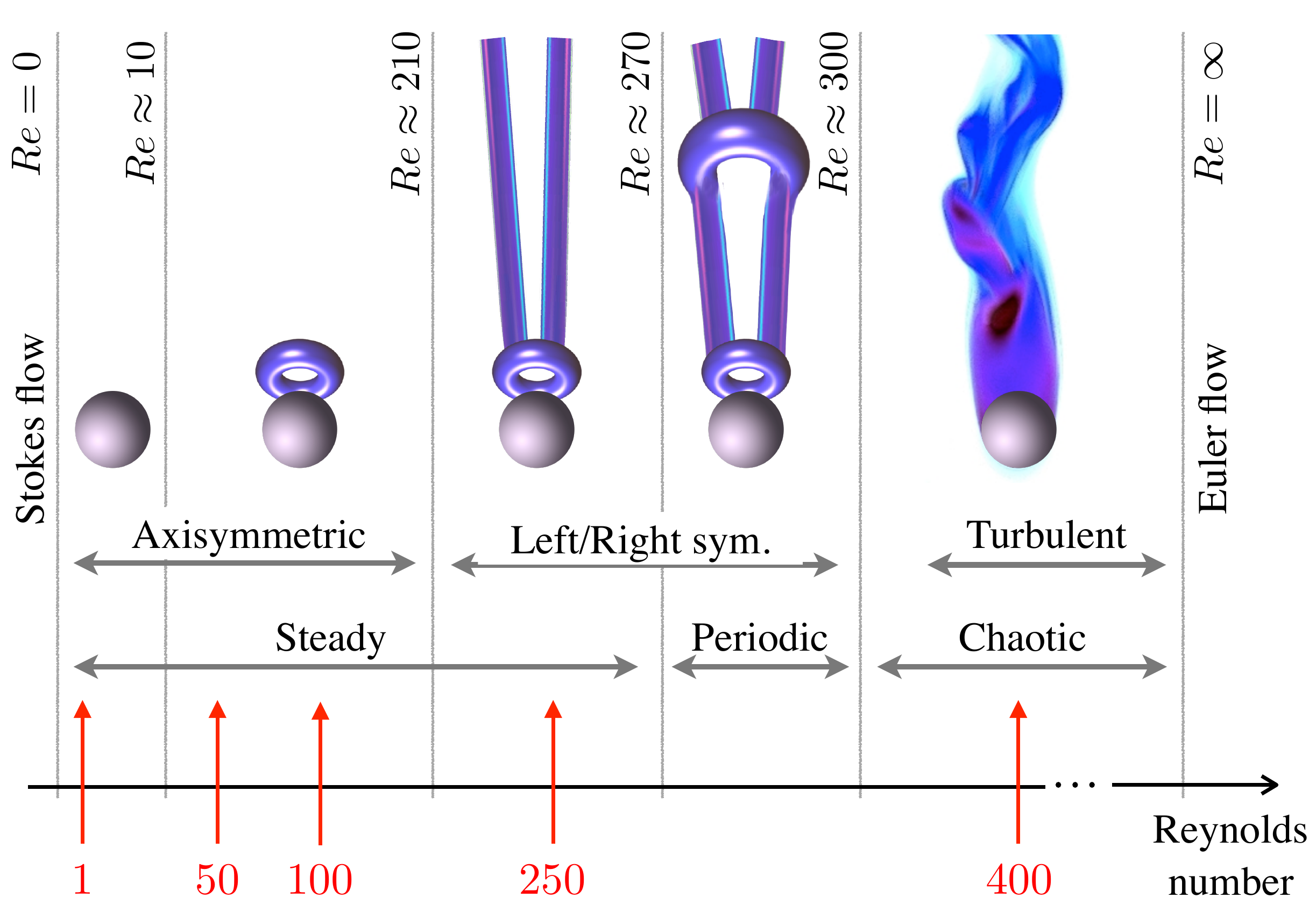}
  \vspace{-20pt}
  \caption{Sketch of the various regimes occurring in the flow past an
    immobile sphere as a function of the Reynolds number
    $Re = U\,d/\nu$. The red arrows correspond to the values
    considered in the numerical simulations of this work.}
  \label{fig:sketch_regimes}
\end{figure}
We have considered different values of the Reynolds number $Re$ which
are representative of the various regimes of the flow past a
sphere. It is indeed known (see, \textit{e.g.}, \cite{drazin2002introduction})
that for $Re\lesssim 10$, the fluid velocity is given by Oseen's flow
and thus resembles Stokes' solution with a tiny fore-and-aft
asymmetry. At larger Reynolds a recirculation zone develops
downstream. The velocity remains axisymmetric up to $Re\approx 210$
and develops a stationary double-threaded wake above this value. For
$Re\gtrsim 270$, vortex shedding starts to occur, the flow becomes
time dependent and is characterized by hairpin structures. It then
encounters a number of bifurcations and symmetry losses when
increasing the Reynolds number~\cite{fabre2008bifurcations} before
becoming chaotic for $Re\gtrsim 300$. All these regimes are sketched
in Fig.~\ref{fig:sketch_regimes}.  To span them, we have considered
$Re = 50$, $100$, $250$ and $400$, and compared them to the ideal
cases of a creeping flow ($Re=0$ obtained from the Stokes equation)
and an inviscid potential flow (of possible relevance upstream the
sphere in the limit $Re\to\infty$).  In these two last flows, explicit
formulas are used for the fluid velocity field~(see, \textit{e.g.},
\cite{falkovich2011fluid}). Details on the numerical simulation
parameters are given in Tab.~\ref{tab:num}.

\begin{table}[h]
  \caption{Parameters of the numerical simulations. All simulations
    were performed with a sphere of diameter $d=1$, in a domain of
    size $50\times 50\times 90$ and a fluid velocity maintained at
    $U=1$ on the upstream and lateral boundaries. The Reynolds number
    is varied by changing the viscosity $\nu$. The total number of
    elements $N_{\rm elem}$, the minimum element size $\delta x_{\rm
      min}$, and the time step, $\delta t$ are varied
    accordingly. \label{tab:num}}
\begin{ruledtabular}
\begin{tabular}{ccccc}
$Re$   & $\nu$       & $N_{\rm elem}$ & $\delta x_{\rm min}$ & $\delta t$ \\
$1$     & $1.0$       & $3.10^6$        & $0.005$                 & $0.002$ \\
$50$   & $0.02$     & $3.10^6$        & $0.005$                 & $0.002$ \\
$100$ & $0.01$     & $3.10^6$        & $0.005$                 & $0.002$ \\
$250$ & $0.004$   & $3.10^6$        & $0.005$                 & $0.002$ \\
$400$ & $0.0025$ & $6.10^6$        & $0.002$                 & $0.001$ \\
\end{tabular}
\end{ruledtabular}
\end{table}

These flows past the sphere are seeded with heavy, inertial,
point-like particles, whose trajectories $\bm X_{\rm p}$ follow
\begin{equation}
  \frac{\mathrm{d}^2\bm X_{\rm p}}{\mathrm{d}t^2} =
  -\frac{1}{\tau_{\rm p}}\left[\frac{\mathrm{d}\bm X_{\rm
        p}}{\mathrm{d}t} - \bm u(\bm X_{\rm p},t)\right].
  \label{eq:particles}
\end{equation}
The particles are assumed to be much smaller than the smallest active
scale of the flow and at the same time sufficiently massive so that
added-mass, Magnus, and history effects can be neglected. They only
interact with the flow through a viscous drag whose intensity is given
by the response time
$\tau_{\rm p} = \rho_{\rm p} d_{\rm p}^2 / (18\,\nu\,\rho_{\rm f})$,
where $\rho_{\rm p}$ and $\rho_{\rm f}$ are the particle and fluid
mass densities, respectively and $d_{\rm p}$ denotes the diameter of
the small particles.  In the numerics, the particles are simulated
using a Lagrangian approach: their trajectories are tracked by
integrating (\ref{eq:particles}) with the fluid velocity at the
particle location obtained by linear interpolation from the
finite-element field. The particles are uniformly injected far
upstream the sphere (at $z=-10\,d$ with a velocity equal to that of
the fluid).  In the steady axisymmetric settings (Stokes and Euler
flows, $Re=1$, $50$ and $100$), the particle dynamics is directly
integrated in the plane $(\rho,z)$ where $\rho^2=x^2+y^2$, instead of
the full three-dimensional space $(x,y,z)$.

\subsection{Collision efficiency}
 \label{sec:coll-eff}

\begin{figure}[h]
  \includegraphics[width=\columnwidth]{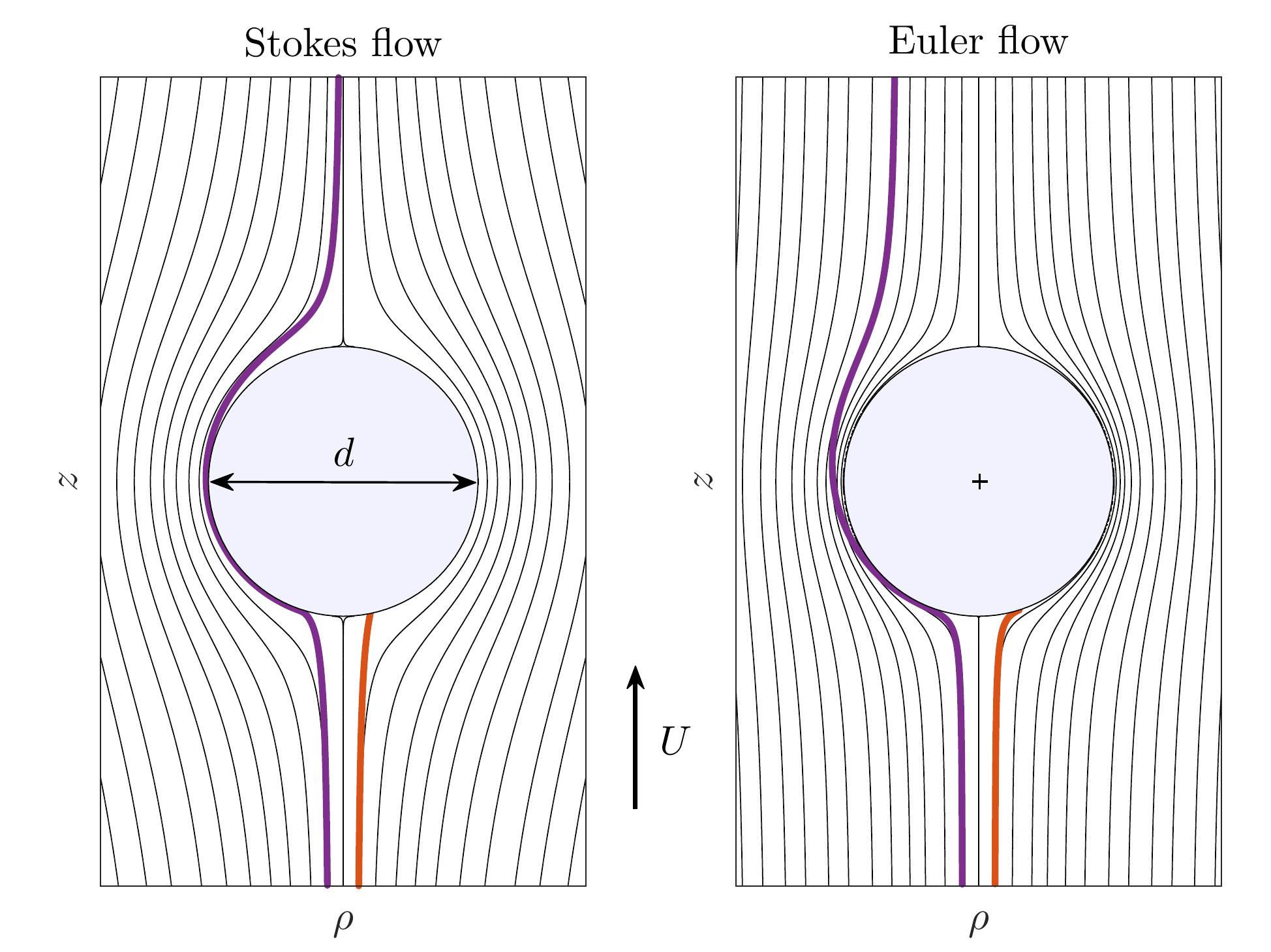}
  \vspace{-20pt}
  \caption{Streamlines corresponding to the two limiting axisymmetric
    cases of a creeping (Stokes) flow (Left) and a potential inviscid (Euler)
    flow (Right). Two particle trajectories have been represented in
    both cases as bold curves.  One which collides with the sphere
    (right-most, in red), and one which does not (left-most, magenta).}
  \label{fig:sketch_stokes_euler}
\end{figure}

The particles that we consider have some inertia whose intensity is
measured by the non-dimensional Stokes number
$St = \tau_{\rm p}\,U/d$.  When $St\to 0$, they are simple tracers and
follow the flow streamlines. Because of the no-penetration condition
at the sphere surface, pointwise tracers never collide with it. The
particles however deviate from the fluid as soon as $St>0$. For
$St\to\infty$, they do not respond anymore to the fluid motion and
simply fly ballistically toward the sphere. These opposite behaviors
exist independently of the specific boundary conditions, being no-slip
as in the case of viscous Stokes flow or free-slip in the case of
inviscid Euler (see Fig.~\ref{fig:sketch_stokes_euler}).
The transition from one behavior at $St\ll1$ to the other at $St\gg1$
is not progressive but there exists a critical Stokes number
$St_{\rm c}$ below which no particles collide with the
sphere. $St_{\rm c}$ is a decreasing function of the Reynolds
number. It is approximately $0.605$ for $Re=0$ (creeping flow) and
approaches asymptotically as $Re\to\infty$ the value $1/24$ obtained
in the potential inviscid case~\cite{phillips1999influence}. The
trajectories represented in Fig.~\ref{fig:sketch_stokes_euler} were
chosen with Stokes numbers below and above the critical values of
these two limiting cases. Note that the reasons why there exists a
critical value of the Stokes number differ whether we consider the
inviscid potential (Euler) flow or viscous settings.  In the former
case, the dynamics is linear in the vicinity of the stagnation point
on the symmetry axis and $St_{\rm c}$ corresponds to a phase
transition where purely real eigenvalues become complex conjugate.
For viscous flows, the velocity normal to the sphere surface vanishes
quadratically and the problem becomes nonlinear. There always exist
colliding trajectories but the Stokes number needs to be large enough
for these trajectories to be physically admissible (\textit{i.e.}\/
recover the fluid velocity at $z=-\infty$). More details can be found
in~\cite{phillips1999influence}.

For Stokes numbers above the critical value $St_{\rm c}$, there is a
beam of particles impacting the sphere around the symmetry axis. In
the limit $St\to\infty$, all particles located upstream in the cross
section of the sphere will collide with it. For finite values of the
Stokes number, some of the particles are deviated by the fluid flow
and escape without impacting. The ratio between the fluxes of
particles that actually collide and of those contained in the swept
volume of the sphere defines $\mathcal{E}_{\rm coll}$, the collision
efficiency. $\mathcal{E}_{\rm coll}=1$ for $St\to\infty$, and
$\mathcal{E}_{\rm coll}=0$ when $St<St_{\rm c}$.

\begin{figure}[h] 
  \includegraphics[width=\columnwidth]{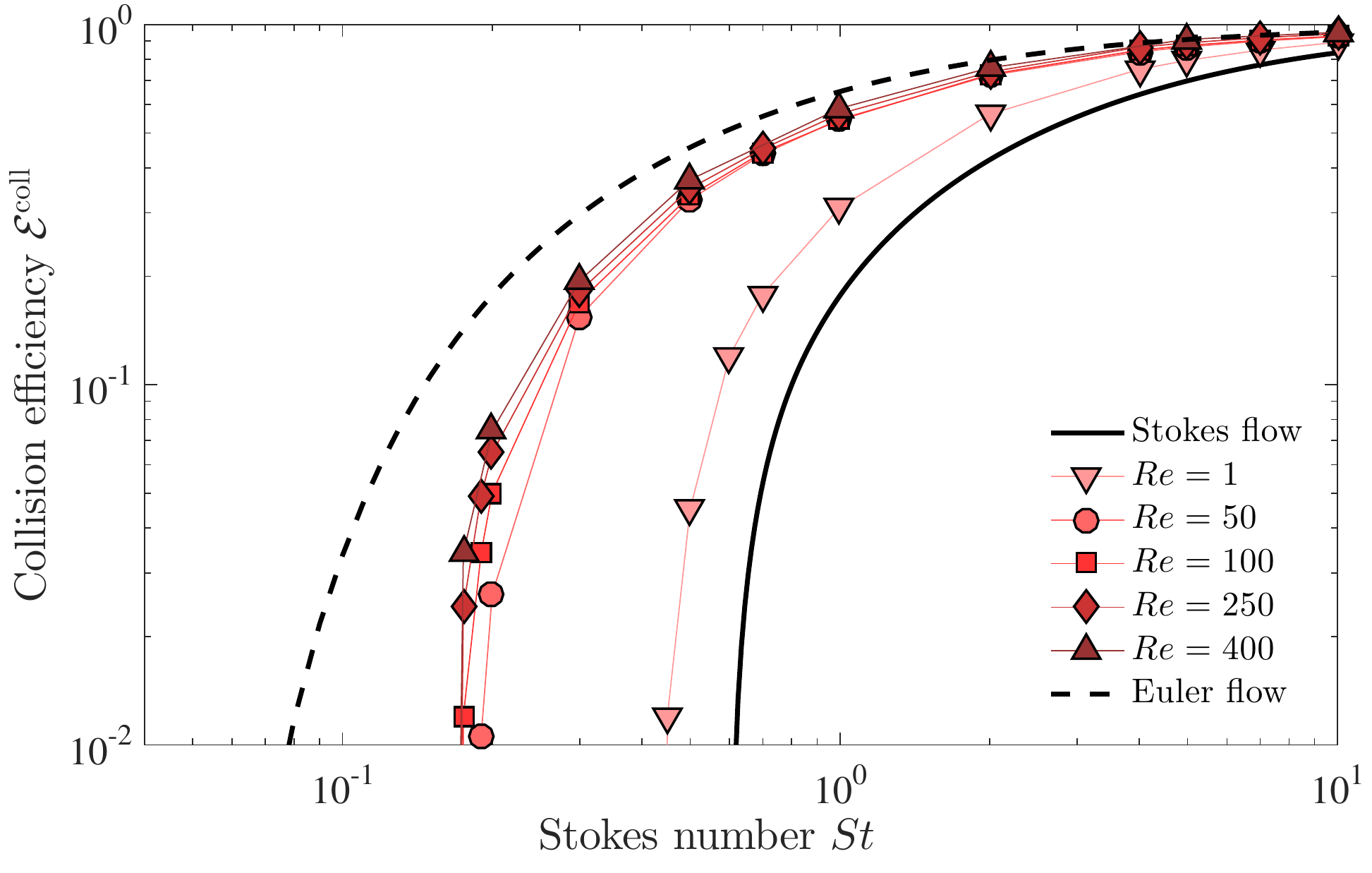}
  \vspace{-20pt}
  \caption{Collision efficiency $\mathcal{E}_{\rm coll}$ as a function
    of $St$ for the various Reynolds numbers considered here,
    including $Re=0$ (Stokes creeping flow) and $Re=\infty$ (Euler
    inviscid flow). }
  \label{fig:Ecoll_St}
\end{figure}
The collision efficiencies measured in our simulations are shown in
Fig.~\ref{fig:Ecoll_St} as a function of the particles Stokes
number. Data corresponding to finite values of the Reynolds number are
embraced by those corresponding to the two limiting ideal cases: the
Stokes creeping flow ($Re=0$) and the Euler inviscid potential flow
($Re=\infty$). One also observes that for a fixed value of the Stokes
number, the efficiency increases as a function of $Re$. This could be
only a consequence of the decrease of $St_{\rm c}$ as a function of
$Re$.  Anyhow, the measured efficiencies remarkably approach the
limiting Euler case.  When the Reynolds number is large, deviations to
the potential flow are restricted to the turbulence that develops in
the sphere wake.  The probability that a particle collides is
essentially determined by the upstream fluid flow which is well
described by the potential case with a small viscous boundary layer of
thickness $\delta_\nu\propto d / \sqrt{Re}$.

Actual values of the collision efficiency are required in the models
used for applications.  Effective and accurate predictions critically
depend upon the values of $\mathcal{E}_{\rm coll}$ that are
prescribed. A frequently used fit is that proposed by Slinn
in~\cite{slinn1983precipitation}, namely
\begin{equation}
  \left\{ \begin{array}{l} \mathcal{E}_{\rm coll} \approx
            \left( \dfrac{St-St_{\rm c}}{St-St_{\rm c}+2/3}\right) ^{3/2}\\
            \\
            St_{\rm c} \approx \dfrac{0.6 + (1/24)
            \log(1+Re/2)}{1+\log(1+Re/2)}.\end{array}
      \right. \label{eq:slinn}
\end{equation}
As already stressed in~\cite{homann2016effect}, such a formula
predicts $\mathcal{E}_{\rm coll} \propto (St-St_{\rm c})^{3/2}$ when
$St\to St_{\rm c}$, which is incompatible with the linear behavior
observed in numerics. We provide here an explanation why one expects
$\mathcal{E}_{\rm coll} \propto (St-St_{\rm c})$ near the critical
Stokes number.

Evaluating the collision efficiency near $St_{\rm c}$ consists in
finding whether or not particles initially located at a distance
$\rho_0\ll d$ from the symmetry axis collide with the sphere. For
that, one expands the fluid velocity near the upstream pole of the
sphere as
\begin{equation}
  u_z \approx a\,z^2+b\,\rho^2, \quad u_\rho \approx -a\,z\,\rho,
\end{equation}
where $a$ and $b$ are positive constants.  We have here assumed that
the flow is viscous and axi-symmetric around the $\rho=0$ axis.  For
particles with Stokes number $St = St_{\rm c} (1+\varepsilon)$ where
$0<\varepsilon\ll 1$, the particle trajectory
$\bm X_{\rm p} = (Z,\rho,\theta)$ in cylindrical coordinates reads to
leading order for axisymmetric settings
\begin{eqnarray}
  \ddot{Z} &\approx&-(1/\tau^{\rm c}_{\rm p})\left[ \dot{Z}
                     -a\,Z^2 -b\,\rho^2-\varepsilon\,\ddot{Z}
                     \right], \label{eq:Zcritical} \\
  \ddot{\rho} &\approx& -(1/\tau^{\rm c}_{\rm p}) \left[\dot{\rho}+
                        a\,Z\,\rho\right], \label{eq:rhocritical}
\end{eqnarray}
where dots denote time derivatives and $\tau_{\rm p}^{\rm c}$ is the
response time corresponding to $St_{\rm c}$. Such a trajectory
corresponds to the critical case of a particle touching the sphere at
infinite time if the two last terms in the right-hand side of
(\ref{eq:Zcritical}) sum to zero. Indeed, in that case $Z$ exactly
follows the same dynamics as the critical trajectory located on the
symmetry axis for $St = St_{\rm c}$.  These two terms cancel out if
$\rho^2$ is of the order of $\varepsilon$. In addition, as the
evolution (\ref{eq:rhocritical}) of $\rho$ is to leading order linear,
one obtains that $\rho_0^2 \sim \rho^2 \sim\varepsilon$, so that
$\mathcal{E}_{\rm coll} = (2\rho_0/d)^2 \propto (St/St_{\rm c}-1)$.

The above considerations are limited to the case of viscous flows
where, as mentioned earlier, the local dynamics does not change
qualitatively at $St=St_{\rm c}$. The situation is different for
inviscid potential flows where a transition occurs at $St_{\rm c}$ and
some timescales diverge. In that case, the dynamics on the symmetry
axis close to the stagnation point reads
\begin{equation}
  \ddot{Z} = -(1/\tau_{\rm p})\left[\dot{Z} + c\,Z\right]
\end{equation}
with $c$ a positive constant ($c = 6\,U/d$ for the Euler flow past a
sphere). This linear system has two eigenvalues. They are real for
$c\,\tau_{\rm p} \le 1/4$ and complex conjugate above. They are then
equal to
\begin{equation}
\lambda = -({1}/{\tau_{\rm p}}) \left[-1\pm i\,\sqrt{4\,c\,\tau_{\rm
      p}-1} \,\right].
\end{equation}
For particle response times right above the critical value, \textit{i.e.}\/
$\tau_{\rm p} = (1+\varepsilon)/4c$, the frequency associated to the
imaginary part of the eigenvalues behave as
$\sqrt{\varepsilon}$. Hence particles released at an order unity
distance from the sphere need a time of the order of
$1/\sqrt{\varepsilon}$ before impacting it. If these particles are at
a distance $\rho_0$ from the symmetry axis, the transverse distance at
impact will be of the order of
$\rho = \rho_0 \exp(C /\sqrt{\varepsilon})$, where $C$ is a positive
constant. This exponential growth is due to the fact that the upstream
flow is divergent in the $\rho$ direction. The impact occurs only if
the particles have not been pushed aside the sphere, that is when
$\rho \lesssim d/2$. This implies that the colliding particles must
satisfy $\rho_0 \lesssim (d/2) \exp(-C /\sqrt{\varepsilon})$. Hence,
in the case of inviscid flows, the collision efficiency behaves as
$\mathcal{E}_{\rm coll} \propto \exp (-C'/\sqrt{St-St_{\rm c}})$ above
the critical Stokes number and is thus increasing much slower than for
viscous flows.

\begin{figure}[h]
  \includegraphics[width=\columnwidth]{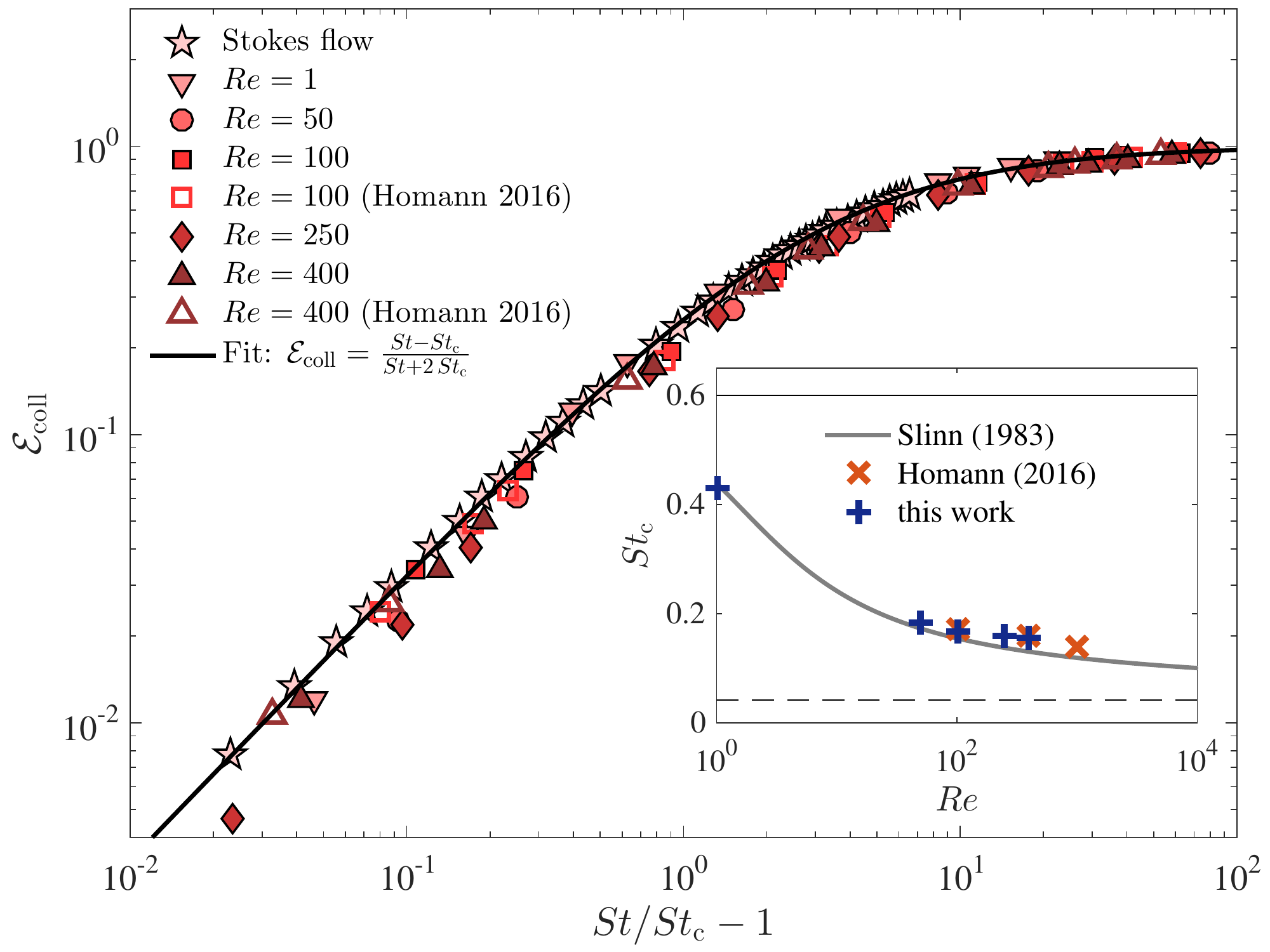}
  \vspace{-20pt}
  \caption{Collision efficiency $\mathcal{E}_{\rm coll}$ as a function
    of $St/St_{\rm c}-1$ where $St_{\rm c}$ is the critical Stokes
    number below which no collisions happen. The data of the various
    simulations are represented as filled symbols, as labeled.  The
    results of Homann \textit{et al.}~\cite{homann2016effect} at
    comparable Reynolds numbers are shown as empty symbols.  The solid
    curve corresponds to the fitting master
    curve~(\ref{eq:fit_new}). Inset: Critical Stokes number as a
    function of the Reynolds number for the various runs.  The
    formula~(\ref{eq:slinn}) proposed by
    Slinn~\cite{slinn1983precipitation} is shown as a solid curve. The
    two horizontal lines are the two asymptotes (Stokes flow
    $St_{\rm c} = 0.605$ and Euler flow $St_{\rm c} = 1/24$).}
  \label{fig:Ecoll_St_over_Stcr}
\end{figure}
These phenomenological arguments show that the behavior of the
collision efficiency is qualitatively different for viscous and
inviscid flows. Thus, for any finite value of the Reynolds number, the
dependence of $\mathcal{E}_{\rm coll} $ on the Stokes number will be
given by the viscous boundary layer dynamics when $St$ is close to
$St_{\rm c}$. The singular behavior of $\mathcal{E}_{\rm coll}$ in
inviscid flows can nevertheless have a signature at very large values
of $Re$ and lead to a deficit of accretion of particles, even if their
Stokes number is well above $St_{\rm c}$. Such an effect could be
responsible for overestimating the actual value of $St_{\rm c}$ in
experiments.  This is not the case in our simulations where we
consider moderate values of the Reynolds number. As seen from
Fig.~\ref{fig:Ecoll_St_over_Stcr}, a linear boundary-layer behavior of
$\mathcal{E}_{\rm coll} $ can clearly be observed for
$St-St_{\rm c}\ll 1$. The boundary-layer considerations detailed above
also suggests that the collision efficiency depends upon
$St/St_{\rm c}$ only. Surprisingly, all data of
Fig.~\ref{fig:Ecoll_St}, including those associated to the Stokes
flow, almost collapse on the top of each other once represented as a
function of $St/St_{\rm c}-1$ in Fig.~\ref{fig:Ecoll_St_over_Stcr}.
An approximation of the master curve is shown as a solid curve. The
fitting was obtained from the data corresponding to the Stokes flow
and reads
\begin{equation}
  \mathcal{E}_{\rm coll} \approx \frac{St-St_{\rm c}}{St + 2\,St_{\rm
      c}}.
  \label{eq:fit_new}
\end{equation}
Figure~\ref{fig:Ecoll_St_over_Stcr} also contains data from Homann
\textit{et al.}~\cite{homann2016effect} which display the same
behavior, except at values of the Stokes number which are maybe too
close to $St_{\rm c}$.  Of course, we have not represented the case of
Euler potential flow which displays a very different behavior.  We
have nevertheless observed that numerical values of
$\mathcal{E}_{\rm coll}$ are in that case compatible with
$\exp (-C'/\sqrt{St-St_{\rm c}})$.  The inset of
Fig.~\ref{fig:Ecoll_St_over_Stcr} represents the measured critical
Stokes number as a function of $Re$. Again, our data are in good
agreement with those of Homann \textit{et al.}~\cite{homann2016effect}
and seem well fitted by Slinn's formula~(\ref{eq:slinn}) for
$St_{\rm c}$.

\subsection{Particles-boundary interactions}
\label{sec:part-bound}

\begin{figure}[h]
\includegraphics[width=.7\columnwidth]{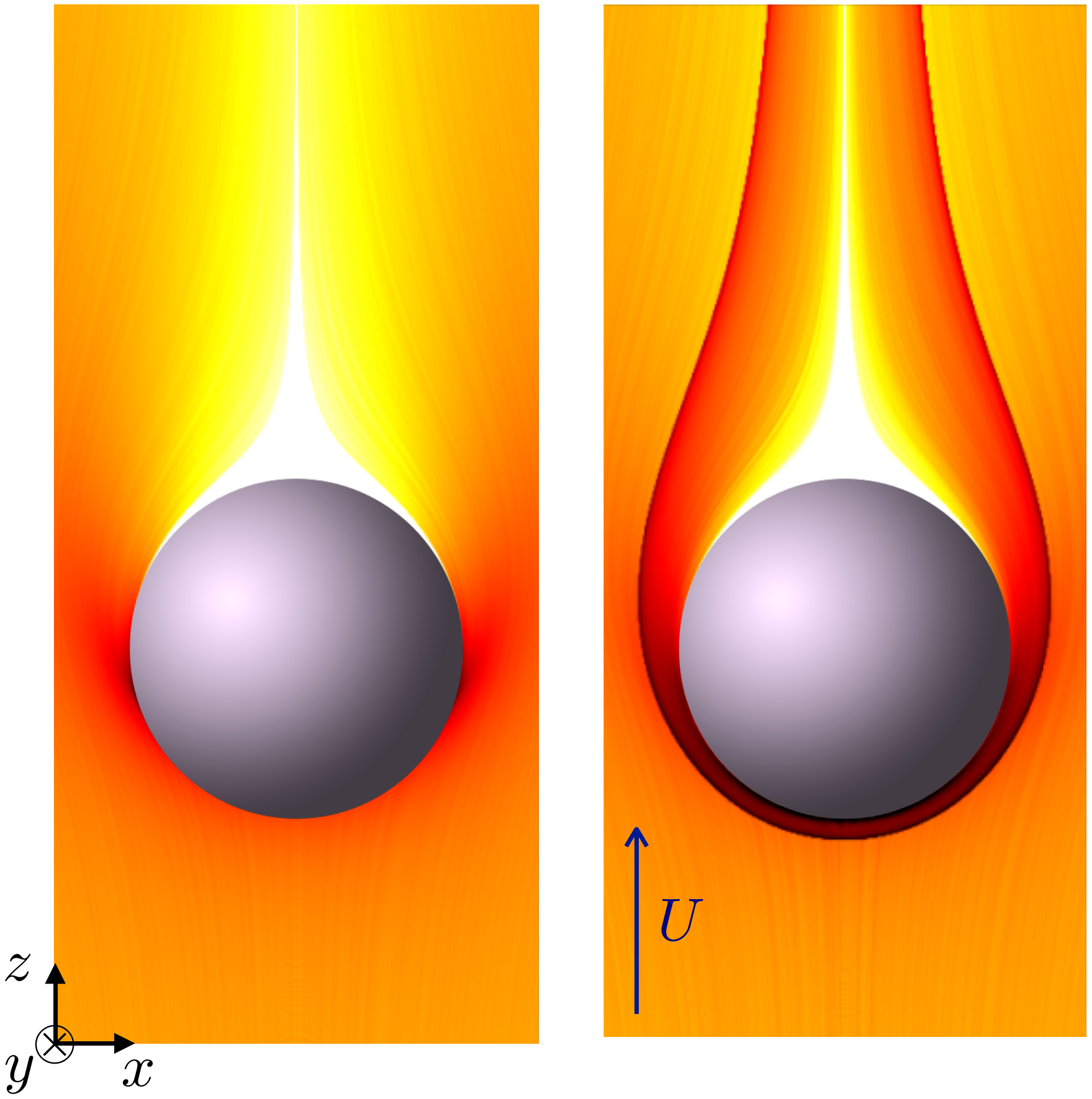}
\caption{Average concentration in the $y=0$ plane of particles with
  Stokes number $St = \tau_{\rm p} U / d = 1$ in a $Re = 0$
  flow with an absorbing boundary condition at the surface of the
  sphere (Left) and when they bounce (Right) with a coefficient of
  restitution $e=0.5$.}
\label{fig:sketch_conc}
\end{figure}

We have discussed in the previous subsection which particles can
collide or not with the sphere as a function of their Stokes number
and of the flow Reynolds number. This has of course important
consequences on accretion. However, the full problem requires
qualifying the outcome of collisions, once they occur.  Depending
whether the particles bounce, break-up, or are absorbed, particle
fluxes at the large object boundary surface can be strongly
altered. This is illustrated in Fig.~\ref{fig:sketch_conc}, which
represents the particle concentration for the same flow and particles
with the same Stokes number but which are absorbed (Left-hand panel)
or inelastically bounce at the surface (Right-hand panel).

We introduce in the following the various possible outcomes of an
impact of a small particle with a larger body and how they depend on
the particle-boundary interactions. The outcome of a collision depends
on a range of parameters and varies with the system considered:
\begin{itemize}
\item In the context of meteorology (raindrop formation), binary
  droplet-droplet interactions in the atmosphere lead to three
  different collision regimes
  \cite{ashgriz1990coalescence,qian1997regimes,saffman1956collision}:
  coalescence, off-center separation and near head-on
  separation. These various regimes depend mainly on various
  parameters: the Weber number $We$ (which measures the relative
  importance of fluid inertia to surface tension), the drop size ratio
  and the impact parameter (which measures the difference between
  head-on and grazing collision). It has been observed that
  coalescence occurs mostly at small $We$ (\textit{i.e.} when the
  velocity is small enough not to break droplets) while breakup is
  significant at higher $We$ especially when the collision is close to
  the head-on or grazing configurations and often lead to the
  apparition of satellite droplets.
\item In the case of spraying processes in combustion, extra collision
  regimes have been identified due to the nature of the droplets
  (chemical composition) \cite{sommerfeld2016modelling}: bouncing can
  take place at very small $We$ while breakup happens also through
  stretching or reflexive separation
  \cite{rabe2010experimental}. These different regimes are similar to
  those observed in the case of jet breakups \cite{eggers2008physics}
  as well as bubble breakup \cite{clift2005bubbles}.
\item In the context of wet deposition
  \cite{pruppacher-klett:1997,mircea2000precipitation,berthet2010scavenging},
  the scavenging of aerosol particles by falling droplets results in
  the 'wet deposition' phenomenon. Studies on the impact between
  droplets and particles have shown that the capture of aerosol by
  falling raindrop takes place under various regimes: Brownian and
  turbulent shear diffusion, inertial impaction, diffusiophoresis,
  thermophoresis and electrical charge effects. Particle washout thus
  depends on a number of parameters related to the fluid flow
  (including the relative humidity) as well as on the nature of the
  aerosols (chemical composition, size, electromagnetic properties)
  \cite{chate2003scavenging}.
\item For collisions between rigid bodies, three regimes have been
  identified \cite{elimelech2013particle,maximova2006environmental}:
  aggregation (where particles stick to each other to form a larger
  aggregate), bouncing or fragmentation (where aggregates breakup in
  several sub-aggregates). The various outcomes depend on several
  parameters, among which: the impact velocity, the impact angle, the
  fluid chemical properties (electrolyte concentration, presence of
  polymers), the particle/aggregate size and geometry.
\end{itemize}

From this brief overview of the possible outcomes of collisions, three
main features can be identified: aggregation (also called capture or
coalescence), bouncing and fragmentation (also referred to as
breakup). In the present case of small inertial particles impacting a
large sphere, we focus mostly on the first two regimes, \textit{i.e.}
capture and bouncing, while the case of fragmentation is left out for
future studies. The distinction between capture and bouncing is
described by rigid body impact theories \cite{stronge2004impact}. A
collision between two rigid bodies occurs in a very brief period of
time during which contact forces prevent particles from
overlapping. Such contact forces (or interface pressure) arise in a
small area of contact between the two bodies and can result in local
deformations. In an elastic collision (typically for hard spheres at
small impact velocities), the contact forces at play are conservative
(\textit{i.e.} reversible) such that the energy of the system is
conserved. In an inelastic collision, the contact forces dissipate
energy (through, \textit{e.g.}, irreversible elasto-plastic
deformations). In addition to these forces, a further friction force
can arise in the tangential direction if the bodies are rough, leading
to sliding motion in the contact area which add further complexity to
the system.

To distinguish between elastic and inelastic collisions, we use the
notion of the restitution coefficient $e$ which is a macroscopic
parameter to characterize the effect of dissipative forces on the
energy before and after collisions (a general definition is available
in \cite{stronge2004impact}). In the following, we assume that
surfaces are perfectly smooth (such that friction forces can be
neglected) allowing to consider a simplified 1D model for collisions
where only the wall-normal component of the particle velocity is
modified (there is no energy loss in the tangential direction). More
precisely, we define a restitution coefficient as the ratio between
the particle wall-normal velocity after collision $v_{\bot}^{+}$ to
the wall-normal velocity before collision $v_{\bot}^{-}$:
\begin{equation}
  e = {v_{\bot}^{+}}/{v_{\bot}^{-}}
\end{equation}
The restitution coefficient $e$ ranges between $0$ (inelastic
collision) and $1$ (purely elastic collision). It should be noted that
the case $e=0$ is actually different from sticky particles since the
normal velocity vanishes while the tangential velocity remains
unaffected (even for a flat boundary). Considerations of refined
restitution coefficients for the outcome of the collision that take
into account friction forces (and thus correlations between
wall-normal and tangential velocities) are left out of the present
paper and will be investigated in future studies.

\begin{figure}[h]
  \includegraphics[width=\columnwidth]{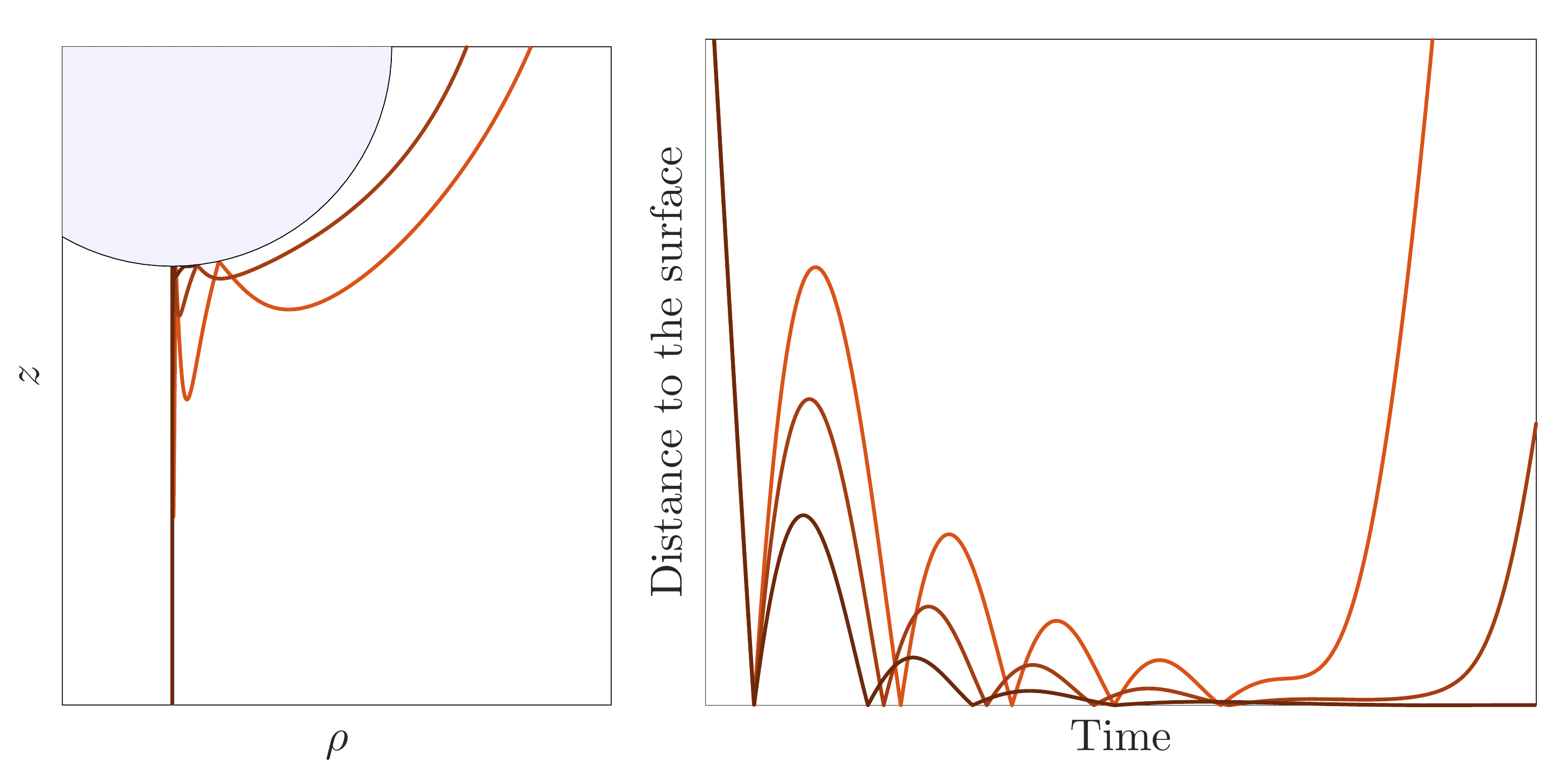}
  \vspace{-20pt}
  \caption{Examples of particle trajectories for $St = 10$ and three
    different restitution coefficients ($e=0.5$, $0.75$ and $1$, dark
    to light red) in the Stokes flow. The particles are injected at
    the same initial value of $\rho$ close to the symmetry axis and
    experience multiple collisions with the sphere. The left panel
    shows their location in the $(\rho,z)$ plane and the right panel
    is the distance to the boundary surface as a function of time.  }
  \label{fig:multiple_coll_traj_stokes}
\end{figure}

Following the analysis of collisions between small particles and a
sphere made in Section~\ref{sec:coll-eff}, we consider inelastic
collisions where the wall-normal velocity after impact is modified
using a single restitution parameter $e$. With these settings one
could expect a kind of 'inelastic collapse' due to a relaxation
through multiple rebounds, which could lead to the particle sticking
to the surface in a finite time. Upstream of the sphere, the flow is
pushing the particle back to the surface after an impact and it
bounces again but with a lower energy. Typical trajectories obtained
in such a case are illustrated in
Fig.~\ref{fig:multiple_coll_traj_stokes} for $St=8$ where we can see
up to 4 or 5 bounces depending on $e$.

As we will see in the next section such a collapse cannot happen in a
finite time in the case of a flow around a sphere, even if $e<1$. Such
a collapse can only happen if the particle is trapped due to
attractive forces with the large sphere at small scales (such as van
der Waals forces, gravitation or electric forces). Considering the
simplest case of a Lennard-Jones potential (which is often used to
approximate the interaction between a pair of neutral atoms or
molecules \cite{israelachvili2015intermolecular}), the interaction
energy of two molecules separated by a distance $r$ is given by:
\begin{equation}
 E_{\rm LJ}(h) = E_{\rm well} \left[ \left({r_m}/{r}\right)^{12} -
   2\left({r_m}/{r}\right)^6 \right]
\end{equation}
where $E_{\rm well}$ is the depth of the potential well and $r_m$ is
the distance at which the potential reaches its minimum. The first
term on the right hand side is repulsive and corresponds to Pauli
repulsion at short ranges (preventing overlap of electron orbitals)
while the second term is attractive and describes long-range van der
Waals force. The resulting interaction between particles (obtained by
integration over the volume of the bodies) is also characterized by a
potential well $E_{\rm stick}$ whose value is proportional to the
particle radius. After bouncing, a particle can thus escape from the
potential well if its kinetic energy after impact is high enough
(otherwise, it will remain trapped in the potential well). To evaluate
whether such a capture happens, the kinetic energy of the particle
after impact is monitored and compared to the potential well
$E_{\rm stick}$.

With respect to this analysis, we have chosen to monitor the velocity
at impact $v^{\rm coll}$ in order to investigate the range of possible
collision outcomes.

\section{Relaxation through successive collisions} 
\label{sec:relax_succ_coll}

As we have seen in the previous section, particles with a given
elasticity can possibly perform multiple successive bounces on the
solid boundary. Here, we investigate whether or not the sequence of
such collisions can lead to an accretion within a finite time. To
simplify the discussion, we suppose that the fluid flow is purely
one-dimensional, normal to the solid surface. In the case of the flow
around the sphere, this is equivalent to focus on particles located on
the symmetry axis. We assume that $\bm u = -c\,z^\alpha\, \bm e_z$
with $c>0$ and $\alpha\ge 0$. An attractive force toward the surface
corresponds to the case $\alpha=0$, while for Euler flow $\alpha = 1$
and for Stokes flow $\alpha = 2$.

Let us consider the dynamics between two successive rebounds. We thus
suppose that a particle with response time $\tau_{\rm p}$ is initially
at the surface $Z_{\rm p}(0) = 0$ with a positive velocity
$\dot{Z}_{\rm p}(0) = v$. For $\alpha>0$, rescaling time by $s = t/T$
with $T= (v/c)^{1/\alpha}/v$ and introducing
$z_{\rm p}(s) = Z_{\rm p}(t) / (v/c)^{1/\alpha}$ leads to
\begin{equation}
  \ddot{z}_{\rm p} = -\frac{1}{S_0} \left(\dot{z}_{\rm p}+{z}_{\rm
      p}^\alpha\right), \quad {z}_{\rm p}(0)=0 \mbox{ and }
  \dot{z}_{\rm p}(0)=1.
  \label{eq:dynam_intercoll}
\end{equation}
The dynamics depends only on $\alpha$ and the Stokes number
$S_0 = \tau_{\rm p}/T = \tau_{\rm
  p}\,c^{1/\alpha}\,v^{1-1/\alpha}$. The problem is then to understand
under which conditions on $\alpha$ and $St$ the particle touches again
the surface $z=0$, and if it does so, at what time and with which
velocity.

Let us first consider the case $\alpha
=1$. Equation~(\ref{eq:dynam_intercoll}) is then linear. One can check that
if $S_0<1/4$, the particle does not go back to the surface but tends
exponentially to it as $t\to\infty$ with a rate
$-(1+\sqrt{1-4\,S_0})/(2\,S_0)$. There is another rebound if
$S_0>1/4$. The solution indeed reads in that case
\begin{equation}
  {z}_{\rm p}(s) = \frac{1}{\omega}{\rm e}^{-s/(2\,S_0)} \sin(
  \omega\,s),\mbox{ with } \omega = \frac{\sqrt{4\,S_0-1}}{2\,S_0}.
  \label{eq:sol_alpha1}
\end{equation}
The rebound is at time $\Delta s = \pi/\omega$ and the impact velocity
is $\dot{z}_{\rm p}(\Delta s) = -\exp(-\pi/\sqrt{4\,S_0-1})$.

\begin{figure}[h]
  \includegraphics[width=\columnwidth]{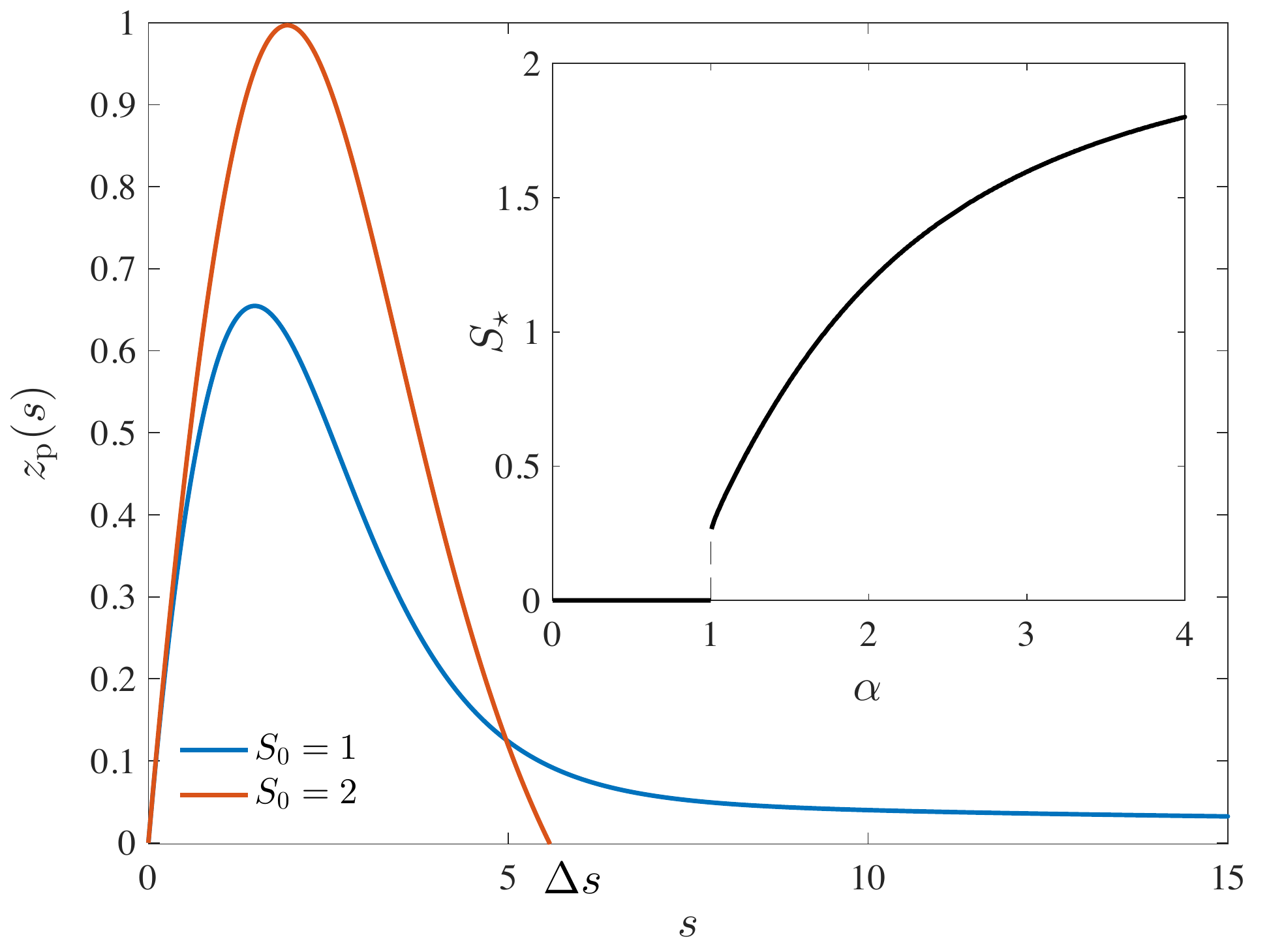}
  \caption{Two trajectories of the system (\ref{eq:dynam_intercoll})
    with $\alpha=2$ associated to values of the Stokes number $S_0$
    below (in blue) and above (in red) the critical Stokes number
    $S_\star(\alpha\!=\!2)\approx1.18$. Inset: critical Stokes number
    $S_\star$ as a function of the fluid velocity exponent $\alpha$.}
  \label{fig:2traj_alpha2}
\end{figure}
\begin{figure}[h]
  \includegraphics[width=\columnwidth]{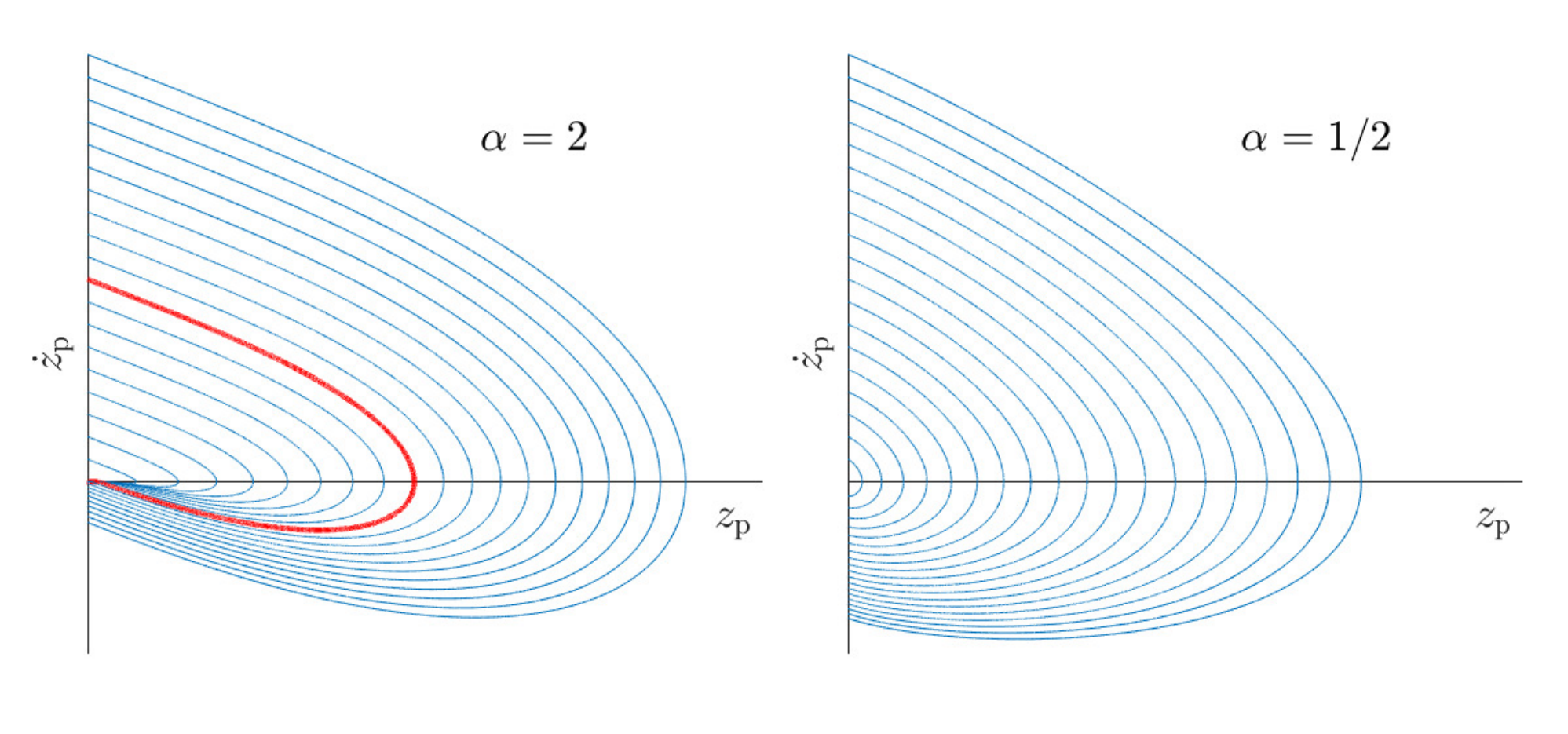}
  \vspace{-20pt}
  \caption{Phase portraits of the $(z_{\rm p},\dot{z}_{\rm p})$
    dynamics for $\alpha=2$ (left) and $\alpha=1/2$ (right). When
    $\alpha>1$ and for any value of the Stokes number $S_0$, there
    exists a critical trajectory (represented as a bold red line) such
    that all trajectories above touch $z_{\rm p}=0$ in a finite time
    with a finite velocity and all those below are tending
    asymptotically to the origin. When $\alpha<1$, there is no such
    critical trajectory and all initial conditions lead to touch
    $z_{\rm p}=0$ in a finite time.}
  \label{fig:phase_portrait2alpha}
\end{figure}

For $\alpha\neq1$ the system is nonlinear and cannot be easily
integrated analytically. Nevertheless, when $\alpha>1$, there still
exists a critical Stokes number $S_\star(\alpha)$ such that the
particles touches again the surface only if $S_0>S_\star$. This is
illustrated in Fig.~\ref{fig:2traj_alpha2} in the case $\alpha = 2$.
When $S_0<S_\star$, the trajectory approaches zero only asymptotically
as a power law ${z}_{\rm p}\propto s^{-1/(\alpha-1)}$. The critical
Stokes number is represented as a function of $\alpha$ in the inset of
Fig.~\ref{fig:2traj_alpha2}. As can be seen from the left-hand side of
Fig.~\ref{fig:phase_portrait2alpha}, $S_0=S_\star$ corresponds to the
case when the critical trajectory (red bold curve) starts from the
initial value $z_{\rm p}=0$ and $\dot{z}_{\rm p}=1$. When $\alpha<1$,
all trajectories hit the surface in a finite time with a finite
velocity, as can be seen from the phase portrait on the right-hand
side of Fig.~\ref{fig:phase_portrait2alpha}.

Hence, all configurations lead to bounce again on the surface if $S_0$
is sufficiently large. The dynamics can be further reduced when
$S_0\gg1$. To leading order the linear damping term in
(\ref{eq:dynam_intercoll}) can indeed be disregarded and
$\ddot{z}_{\rm p} \approx - z_{\rm p}^\alpha/S_0$ with $z_{\rm p}(0)=0$
and $\dot{z}_{\rm p}(0)=1$.  Then, rescaling both time and space by
$S_0^{1/(\alpha+1)}$, one gets rid of the dependence on $S_0$. This
implies that when $S_0\gg1$, the time to next collision scales as
$\Delta s \propto S_0^{1/(\alpha+1)}$.  In this approach, we have
neglected the time-irreversibility of the particle dynamics.
Consequently, the impact velocity is to first order $\dot{z}_{\rm
  p}(\Delta s) \approx -1$. Next order corrections are obtained by
looking at the kinetic energy budget
\begin{equation}
  \frac{1}{2}\dot{z}^2_{\rm
    p}(\Delta s)-  \frac{1}{2}\dot{z}^2_{\rm
    p}(0) = -\frac{1}{S_0}\int_0^{\Delta s} \left[\dot{z}^2_{\rm
      p}+\dot{z}_{\rm p}\,z_{\rm p}^\alpha \right] \mathrm{d}s.
\end{equation}
The second term in the integral, which corresponds to the work done by
the fluid velocity, is to leading order time symmetric and thus
integrates to zero. The main contribution to the energy dissipation is
thus coming from the first term and corresponds to the viscous damping
along the time-reversible trajectory. As $\dot{z}_{\rm p} = O(1)$, the
energy loss between two rebounds is then of the order of
$\Delta s/S_0 \sim S_0^{-\alpha/(\alpha+1)}$.

We now turn to reinterpreting these results in terms of successive
rebounds experienced by a given particle.  Let us assume that the
particle undergoes a sequence of bounces on the solid surface at times
$t_n$ where $n$ indexes the $n$-th collision. At $t=t_n^+$, the
particle leaves the surface with a velocity denoted $v_n$. As seen
above, the dynamics up to time $t_{n+1} = t_n +\Delta t_n$ depends
solely on $\alpha$ and on the Stokes number
$S_n = \tau_{\rm p}\,c^{1/\alpha} \,v_n^{1-1/\alpha}$.  As the
particle looses kinetic energy at each bounce, we expect $v_n$ to be
decreasing sequentially. This implies that $S_n$ increases as a
function of $n$ if $\alpha<1$ and decreases if $\alpha>1$.  In this
latter case, we have also seen above that there exists a critical
Stokes number below which the particle does not touch again the
surface within a finite time. Consequently any collision sequence is
finite when $\alpha >1$.  Any particle that is thrown toward the
surface is experiencing a finite number of collisions until it has
dissipated enough energy. It then converges only asymptotically (as a
power law with exponent $-1/(\alpha -1)$) towards $z=0$ and there is
no accretion within a finite time.  The dynamics between two bounces
roughly consists of two stages: The first one is characterized by a
dissipation of the particle kinetic energy when it goes away from the
solid boundary; The second step consists then in an entrainment of the
particle by the fluid flow toward the surface.  When $\alpha>1$, the
fluid velocity decreases too fast when $z\to0$ and the first step
prevails.  Notice that the effect of non-elastic collisions
(\textit{i.e.}\/ a restitution coefficient $e<1$) increases
dissipation and thus accentuates this phenomenon.

The situation is different when $0<\alpha<1$. The Stokes number $S_n$
increases as a function of $n$, so that dissipation becomes less and
less important.  Any particle that is thrown toward the surface
bounces and experiences an infinite sequence of collisions. After
sufficiently many rebounds, $S_n$ is large enough to apply the
asymptotic results discussed above. The time between the $n$-th and
$(n+1)$-th collisions reads
$\Delta t_n \sim T\,\Delta s \sim v_n^{(1-\alpha)/(1+\alpha)}$.  The
sequence of impact velocities satisfies
\begin{eqnarray}
 v_{n+1} &=& -e\,v_n\,\dot{z}_{\rm p}(\Delta s) \simeq e\,v_n\,\left
   (1-C\,S_n^{-\frac{\alpha}{1+\alpha}} \right) \nonumber \\
  &\simeq& e\,v_n\,\left
   (1-C\,\tau_{\rm p}^{-\frac{\alpha}{1+\alpha}}c^{-\frac{1}{1+\alpha}}v_n^{\frac{1-\alpha}{1+\alpha}}
           \right), \nonumber
\end{eqnarray}
where $e$ is the restitution coefficient of the particles on the
surface and $C$ is a positive constant. This
recurrence relation gives for $n$ large
\begin{equation}
v_n \sim \left\{ \begin{array}{ll} n^{-\frac{1+\alpha}{1-\alpha}}
                   &\mbox{ for } e=1,\\
                   e^n &\mbox{ for } e<1.\end{array} \right.
\end{equation}
For $e=1$, this behavior implies that $\Delta t_n \sim 1/n$ and thus
$t_n$ diverges. The collision sequence lasts forever and an infinite
time is required for the particle to stick to the surface. Conversely,
if $e<1$, the inter-collision time $\Delta t_n$ tends exponentially to
zero and its series converges to a finite time $t_\star$ at which the
particle is accreted at the surface. This closes the case
$0<\alpha<1$, the two extreme values needing to be treated separately.

For $\alpha=0$, which corresponds to a constant attractive force
between the particle and the solid surface, one cannot proceed in
terms of $S_n$ as above but the system can be explicitly integrated,
leading for $n$ large to $\Delta t_n \simeq 2\tau_{\rm p} v_n/c$ and
$v_{n+1} \simeq e\,v_n(1- 4v_n/(3c))$. One thus obtains exactly the same
long-term behavior as for $0<\alpha<1$.

The case $\alpha=1$ corresponds to a free slip boundary condition for
the fluid velocity at the solid surface. As we have seen previously,
the Stokes number reads $S_n = c\,\tau_{\rm p} = S_0$ and is then
independent of $v_n$. If $S_0>1/4$, the time between successive
collisions is constant $\Delta t_n = \pi/(c\,\omega)$, where
$\omega(S_0)$ is given in (\ref{eq:sol_alpha1}). We thus have
$t_n\propto n$.  The kinetic energy is linearly damped between
successive collisions and thus converges exponentially to $0$ as
\begin{equation}
  v_n\propto\exp\left[n\left(\log e
      -{\pi}/{\sqrt{4S_0-1}}\right)\right].
  \label{eq:decreaselinear}
\end{equation}
The effect of inelasticity is just to accelerate the exponential
decrease. The particle converges exponentially fast to the boundary
and there is no accretion within a finite time.

To end this section, let us shortly summarize our findings. A
surprising result is that the only instance when there is accretion in
a finite time by successive bounces requires two conditions to be satisfied:
the collisions should be inelastic ($e<1$) and the fluid velocity should scales
as $u(z) \propto -z^\alpha$ with $\alpha <1$. In the cases of physical
relevance to the accretion by a sphere, we have $\alpha\ge1$ and the
particles are only approaching asymptotically the surface,
exponentially fast for $\alpha=1$ and as a power law for
$\alpha>1$. Hence, the mechanism leading to particle accretion cannot
be related to any kind of inelastic collapse. It is thus needed to
introduce other criterions to determine whether or not particles are
sticking to the solid surface. As motivated in
Sec.~\ref{sec:part-bound}, a relevant quantity is the kinetic
energy upon impact (or equivalently the modulus of the impact
velocity) whose statistics are clearly given by the sequence $v_n$ and
discussed in next section for the problem of the flow past a sphere.
 
\section{Impact statistics}
 \label{sec:impact_stats}

We now turn to describing impact velocities statistics obtained from
the numerical simulations of the flow past a sphere at the various
values of the Reynolds number mentioned in
Sec.~\ref{sec:accretion_problem}.  We have seen in the previous section
that particles cannot approach the sphere with a vanishing velocity
within a finite time through an inelastic collapse.  Thus, the
possibility that some particles stick to the sphere is necessarily
related to the introduction of a finite critical impact velocity below
which accretion occurs (see Sec.~\ref{sec:part-bound}).  This
motivates studying the statistics of the modulus
$v^{\rm coll} = |\mathrm{d}\bm X_{\rm p}/\mathrm{d}t|$ of the particles velocity
at impact as well as its average value $v = \langle v^{\rm coll} \rangle$. 
The idea is to understand how inelastic rebounds influence
the distribution of $v$. As we have seen in previous section,
particles can indeed undergo successive collisions and the impact
velocity $v_n$ at the $n$-th bounce can be possibly related to that at
the previous collision. For that reason, we start with the statistics
of the velocity $v^{\rm coll}_1$ at the first collision.

\begin{figure}[h]
  \includegraphics[width=\columnwidth]{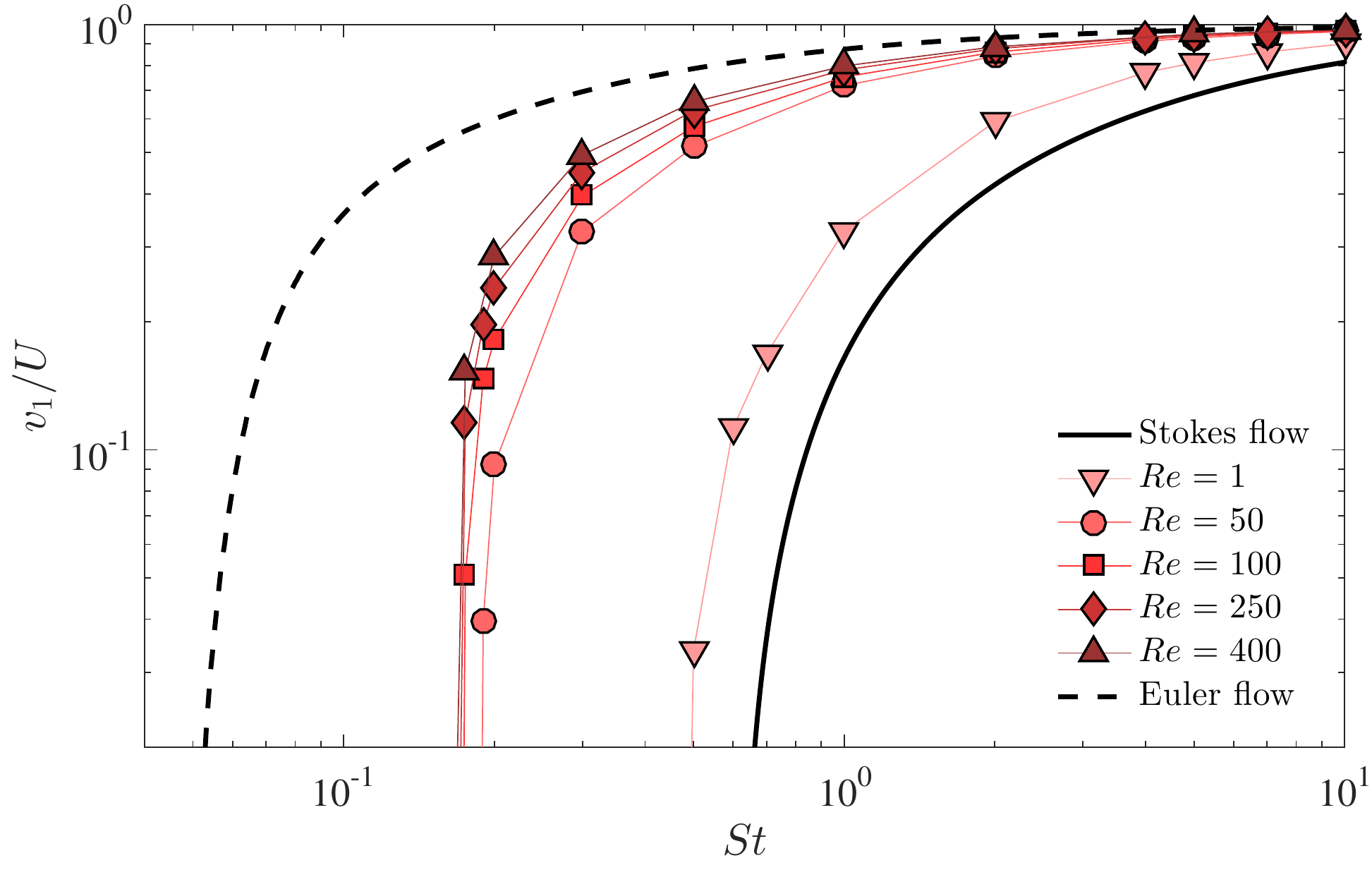}
  \vspace{-20pt}
  \caption{Average of the modulus of the impact velocity
    $v_1 = \langle v_1^{\rm coll}\rangle$ at the first collision, as a function of the
    Stokes number for the various Reynolds numbers, including the
    limiting cases of the Stokes creeping flow and the inviscid
    potential flow.}
  \label{fig:AvgCollEnergy_fn_St_variousRe}
\end{figure}
Figure~\ref{fig:AvgCollEnergy_fn_St_variousRe} shows the average
$v_1 = \langle v_1^{\rm coll}\rangle$ $v_1^{\rm coll}$ of the velocity
modulus at first impact over all particles that collide at least once
with the sphere. Again, as for the collision efficiencies, all data
are contained in between the two limiting cases of the Stokes and
Euler flows.  Despite the fact that the Reynolds numbers span
different regimes of the flow past a sphere, there is no evidence of
any abrupt change in the average impact velocity.  Colliding particles
are indeed only influenced by the upstream flow, which actually does
not undergo any discontinuous changes when $Re$ increases. In
addition, the curves are ordered: For a fixed Stokes number, the
average impact velocity grows as a function of the Reynolds number and
becomes closer to the potential flow.  This can be explained by the
fact that the viscous boundary layer becomes thinner when $Re$
increases, and this is the place where particles are the most
efficiently slowed down. The convergence to the inviscid case is
faster at larger Stokes numbers. In that limit, the effect of the
boundary layer is indeed becoming weaker because the time during which
it affects particles dynamics becomes less than their response time.
Note that there is a wide gap in impact velocity amplitude when moving
from Stokes to Euler flow.  This important increase is due to the fact
that the tangential component of the velocity does not vanish in the
inviscid case.

\begin{figure}[h]
  \includegraphics[width=\columnwidth]{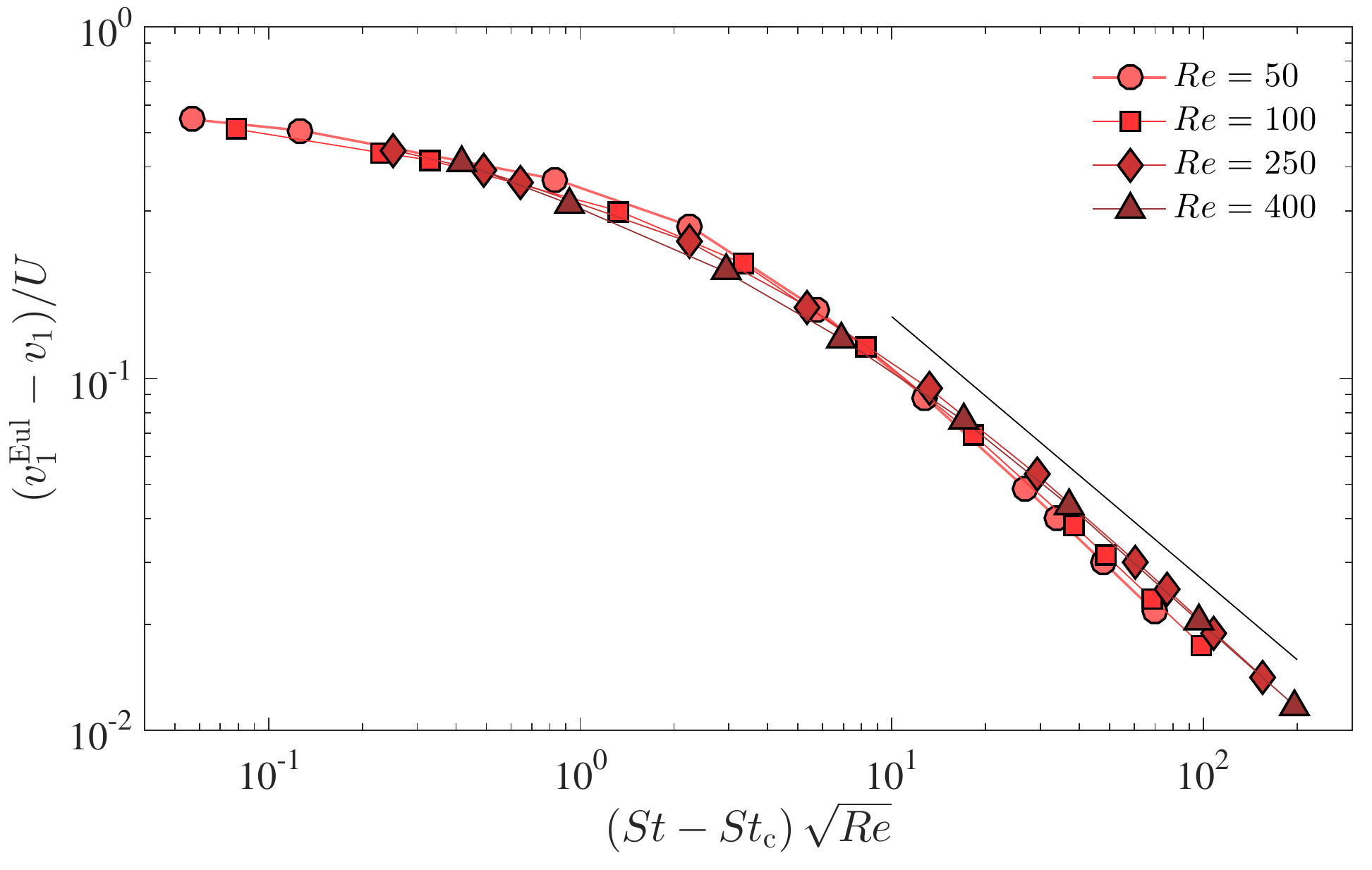}
  \vspace{-20pt}
  \caption{Deviations from the inviscid potential case of the average
    impact velocity at the first collision. They collapse on the top
    of each other when represented as a function of
    $(St-St_c)\,\sqrt{Re}$. The black line shows a slope $-3/4$.}
  \label{fig:Vcoll_Euler_deviation}
\end{figure}
Figure~\ref{fig:Vcoll_Euler_deviation} represents the
deviation from the Euler inviscid case of the average impact velocity
as a function of the Stokes number. We indeed observe that the
deviations vanish when the Stokes number increases.  This can be
understood with the following argument. Let us suppose that a particle
is entering the viscous boundary layer, \textit{i.e.}\/ is at a distance
$\delta_\nu\propto d/\sqrt{Re}$ from the surface, with a radial
velocity $v_1^{\rm Eul}$ that is resulting from the action of the
inviscid Euler flow. For a sufficiently large Reynolds number, we
assume that the fluid velocity seen by the particle is indeed that of
the Euler potential flow until it reaches a distance $\delta_\nu$ from
the sphere. For large Stokes numbers, the fluid velocity in the
boundary layer is always much smaller that the particle velocity. The
particle is thus decelerated as if it was in a fluid at rest and
impacts the surface with a velocity
$v_1 \simeq v_1^{\rm Eul} - \delta_\nu / \tau_{\rm p}$.  The deviation
to the Euler flow is thus indeed decreasing with the Stokes
number. This phenomenology also suggests that the deviations are just
a function of $St \sqrt{Re} = U \tau_{\rm p}/\delta_\nu$. Such a
scaling is indeed confirmed in
Fig.~\ref{fig:AvgCollEnergy_fn_St_variousRe} where representing the
average impact velocity as a function of $(St-St_{\rm c})\sqrt{Re}$
allows one to collapse the results associated to the various Reynolds
numbers. Surprisingly, such a scaling seems to reasonably extend to
the moderate values of the Stokes and/or Reynolds numbers ($Re=1$ has 
a similar trend as the one for Stokes flow and is thus not shown). At large
values, the deviations to the Euler case decreases approximatively as
$v_1^{\rm Eul} - v_1 \sim (St
\sqrt{Re})^{-3/4}$. Such a behavior, which is different from that
obtained above with one-dimensional phenomenological arguments, is
certainly due to the spherical geometry of the boundary layer.

\begin{figure}[h]
  \includegraphics[width=\columnwidth]{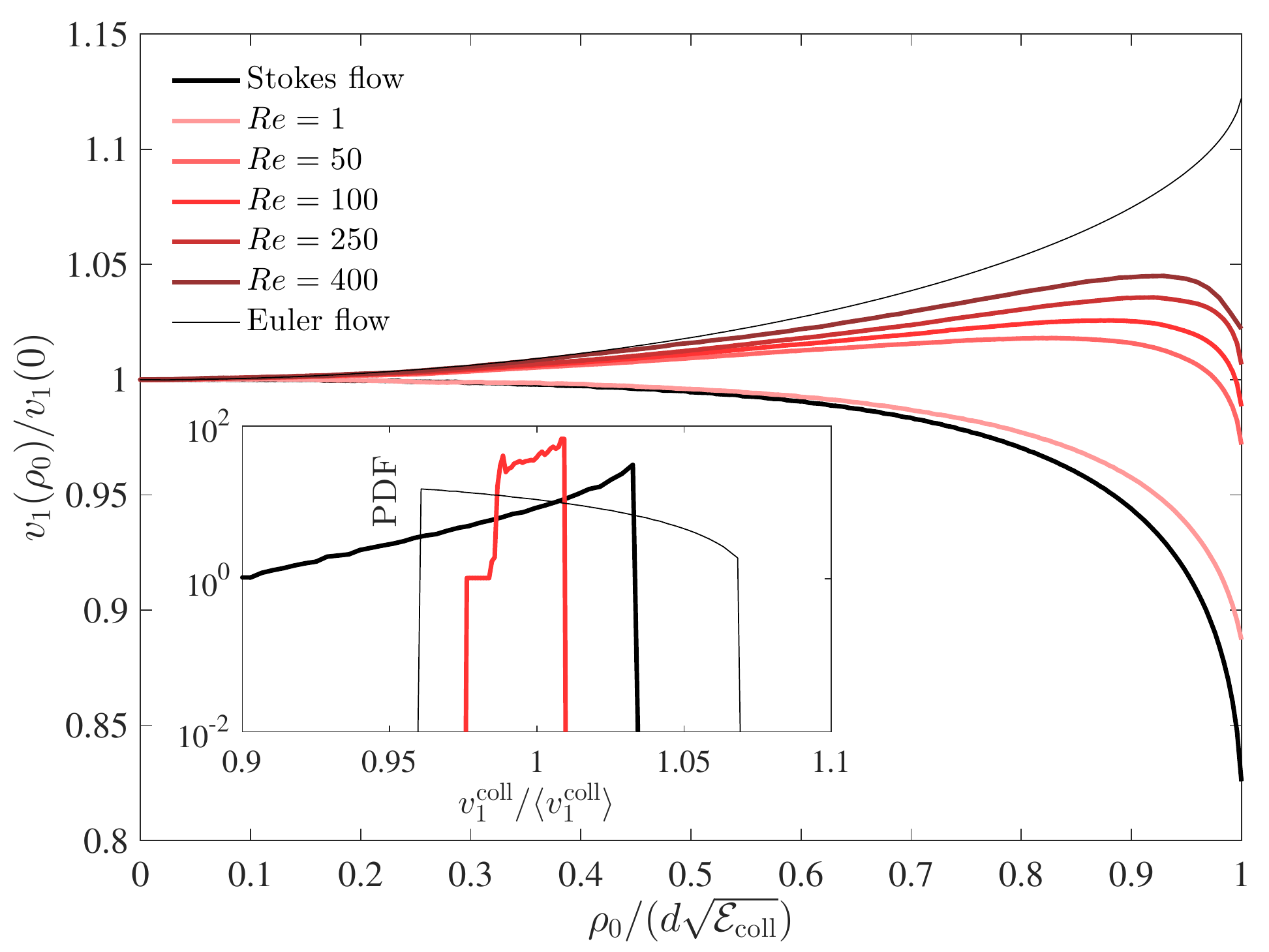} 
  \vspace{-20pt}
  \caption{$v_1^{\rm coll}$ at the first collision as a function of
    the injection location $\rho_0$ for $St=2$ and the
    various Reynolds number that we have considered. 
    Inset: Probability density function of the first impact velocity
    $v_1^{\rm coll}$ for $St=2$ and various flows (Stokes 
    flow, $Re=100$, and Euler flow).}
  \label{fig:CollEnergy_fn_rho_variousRe}
\end{figure}
The behavior of the average impact velocity seems rather simple, at
least for the first collision, and displays qualitative features that
resemble much those of the collision efficiency. This could lead to
postulate that the more likely the collisions are, the more energetic
they are. However, the actual outcomes of collisions cannot be
inferred from such an average: They depend on whether or not the
individual collision energies of each particle are above or below a
critical value. This leads to studying how the impact velocity varies
among the colliding
particles. Figure~\ref{fig:CollEnergy_fn_rho_variousRe} shows the
average value of
$v_1(\rho_0) = \langle v_1^{\rm coll}\,|\, \rho_0\rangle$ conditioned
on the initial distance $\rho_0$ of the particle from the symmetry
axis.  It is here represented for $St=2$ and the various values of the
Reynolds number that we have considered.  One observes that the two
limiting cases of Stokes creeping flow and Euler potential flow
display very different qualitative behaviors. In Stokes flow
$v_1(\rho_0)$ attains its maximum on the symmetry axis and decreases
with $\rho_0$ while in Euler flow the potential flow attains a minimum
at $\rho_0=0$ and then increases. These two different behaviors are
related to the very different nature of the near-surface dynamics. In
particular, the impact velocity vanishes at
$\rho_0 = (d/2)\,\sqrt{\mathcal{E}_{\rm coll}}$ in the Stokes
case. This value corresponds to the colliding particles that are the
furthest from the axis of symmetry. As we have seen in
Sec.~\ref{sec:coll-eff}, this corresponds to the initial conditions of
the critical trajectory that delimits colliding from non-colliding
particles in a quadratic viscous boundary layer (see, \textit{e.g.},
the red bold trajectory on Fig.~\ref{fig:phase_portrait2alpha}).  Such
a trajectory is actually touching the sphere surface but only after an
infinite time and with a vanishing velocity. Such a trajectory does
not exist in the potential flow which vanishes linearly at the sphere
surface, as long as $St>St_{\rm c} = 1/24$. In this case the impact
kinetic energy is essentially given by the amount of deceleration
experienced by the particles before touching. The structure of Euler's
potential flow is such that the $z$ component of the fluid velocity at
a distance of the order of $d/2$ from the symmetric axis is higher
than $U$. Thus the most eccentric particles are those which are the
less slowed down, explaining the increase of $v_1(\rho_0)$ as a
function of $\rho_0$. Finite values of the Reynolds number yield an
intermediate behavior: The impact velocity first increases when moving
away from the symmetry axis, as in the potential case, and vanishes
for $\rho_0 = (d/2)\,\sqrt{\mathcal{E}_{\rm coll}}$, under the effect
of the viscous boundary layer. It attains a maximum in between these
two extremes.

The conditional average $v_1(\rho_0)$ leads to an expression for the
distribution of the impact velocities if we assume that the particles
are homogeneously distributed far upstream. We can indeed write
\begin{equation}
  p(v_1) =
  p(\rho_0)\,\left|\frac{\mathrm{d}\rho_0}{\mathrm{d}v_1}\right| =
  \frac{1}{\mathcal{E}_{\rm
      coll}\,(d/2)^2}\,\left|\frac{\mathrm{d}\rho_0^2}{\mathrm{d}v_1}\right|.
  \label{eq:p_v-p_rho}
\end{equation}
The corresponding distributions are shown in the inset of
Fig.~\ref{fig:CollEnergy_fn_rho_variousRe}. One recovers the two
qualitatively different cases of the creeping and potential flows. The
distribution in Euler flow is rather flat and bends down at the
largest values of $v_1$. Conversely the Stokes flow leads to a peaked
distribution associated to the flat behavior of $v_1$ close to
$\rho_0=0$.  For intermediate Reynolds numbers, the distribution is
obtained as a superposition of these two behaviors. The peak persists
at the maximal value of $v_1$, which is now attained at a finite
$\rho_0$ rather than on the symmetry axis. Also, the distribution
contains two steps associated to the two branches of
$v_1 \mapsto \rho_0$.

\begin{figure}[h]
  \includegraphics[width=\columnwidth]{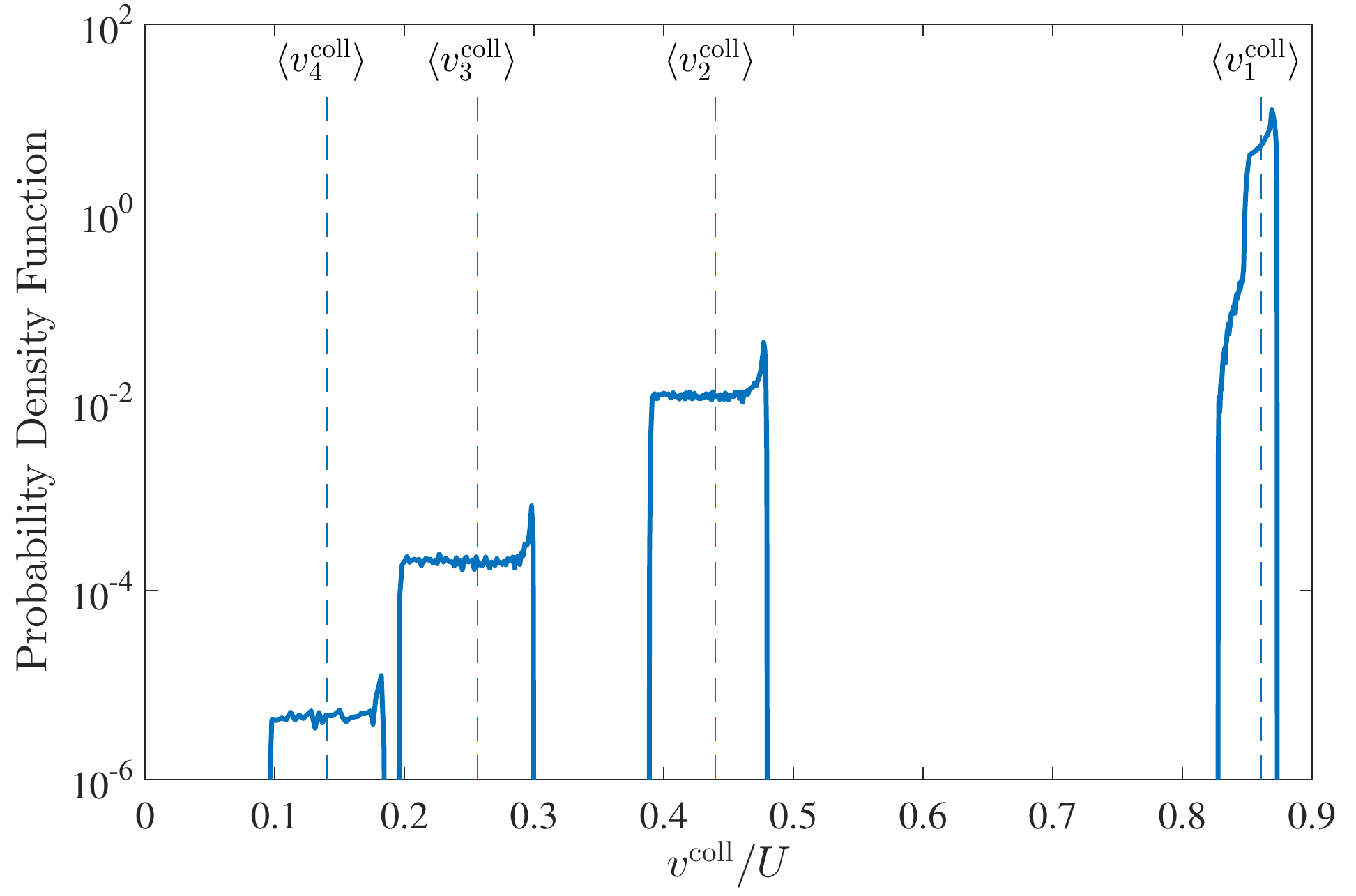}
  \vspace{-20pt}
  \caption{Probability density function of the impact velocity
    $v^{\rm coll}$ for $Re =100$, $St=2$ and $e=1$. The distribution
    is multimodal, with the various modes associated to successive
    bounces of the particles. The vertical dashed lines show the
    averages impact velocities $\langle v_n^{\rm coll} \rangle$ at the
    $n$-th collision.}
  \label{fig:PDFCollEnergy_variousRe}
\end{figure}
\begin{figure}[h]
  \includegraphics[width=\columnwidth]{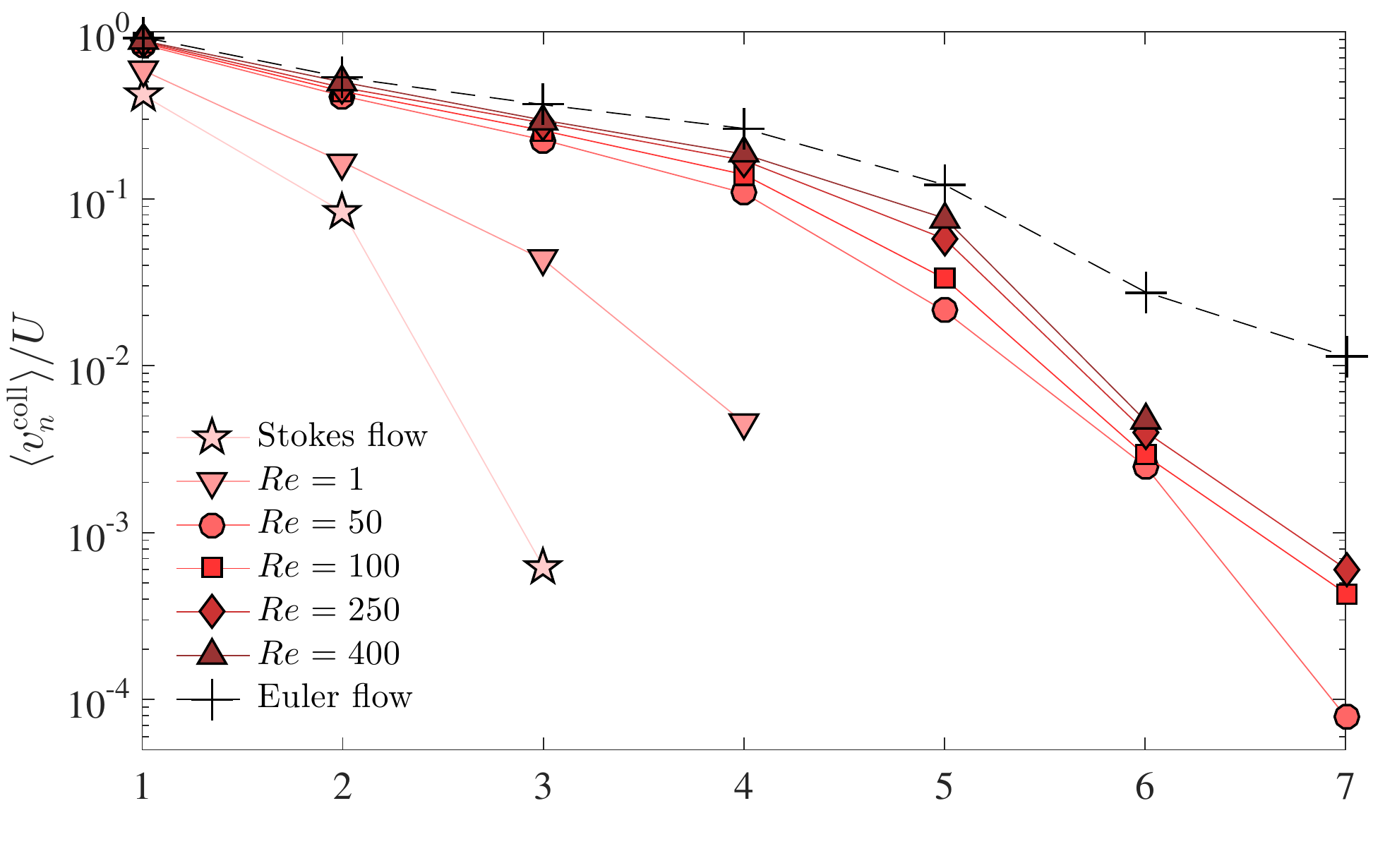}
  \vspace{-20pt}
  \caption{Average of the impact velocity $\langle v^{\rm
      coll}_n\rangle$ at the $n$-th bounce as a function of $n$ for
    $St=2$, $e=1$ and the various Reynolds number that we have
    considered.}
  \label{fig:AvgCollEnergy_fn_n_variousRe}
\end{figure}

Now that we have described impact velocity statistics at the first
collision, we extend here this approach to the case of bouncing
particles. Figure~\ref{fig:PDFCollEnergy_variousRe} shows the
probability density function of the impact velocity $v^{\rm coll}$ for
particles with $St=2$, a coefficient of restitution $e=1$ in the
$Re=100$ flow. One clearly observes a multimodal distribution. The
right-most bump corresponds to the first impact and has a shape as
given above when considering the conditional average of
$v_1^{\rm coll}$. The other modes to the left are associated to the
successive rebounds of the particles. Their shapes are similar to each
other but do not exactly reproduce the first-collision distribution.
All modes are characterized by a flat region at small values
associated to a quadratic minimum on the symmetry axis, and a peak at
large values due to the maximum of impact velocity.  When successive
bounces occur, two phenomena are at play in order to predict the shape
of the next mode in the distribution of $v^{\rm coll}$. The first is a
decrease of the typical value of $v^{\rm coll}$, which might be
described by the results of previous section. The second relates to
the fact that only a fraction of bouncing particles will touch again
the sphere at a later time. This diminishes the probability weight of
each mode when $v^{\rm coll}$ decreases. The combination of these two
effects results in an enlargement of each mode when the number of
bounces increases. In some cases other than the one shown in
Fig.~\ref{fig:PDFCollEnergy_variousRe}, different modes can even
superpose. Except that, all results associated to other values of
$St$, $e$ and $Re$ display similar behaviors. To complete the picture,
we have represented in Fig.~\ref{fig:AvgCollEnergy_fn_n_variousRe} the
average impact velocity $\langle v^{\rm coll}_n \rangle$ at the $n$-th
impact. When the Reynolds number is large enough, two behaviors are
visible as a function of $n$: $\langle v^{\rm coll}_n \rangle$ starts
with decreasing exponentially during the first collisions. This
corresponds to bounces with a large enough energy such that the
particles exit again the viscous boundary layer and the dynamics is
close to that given in Eq.~(\ref{eq:decreaselinear}). Once the
collisions are not sufficiently energetic, the particles are trapped
in the viscous boundary layer and the impact velocity decreases faster
than an exponential.

To complete the picture, we finally propose a way to interpret
previous results in terms of an accretion efficiency. For that
purpose, we focus on the case of the Stokes creeping flow which, as
long as near-boundary dynamics are concerned, is qualitatively
representative of flows developing a viscous layer and is easily
amenable for a detailed systematic study. As explained in
Sec.~\ref{sec:part-bound}, we consider here the simplest model for
accretion: A particle sticks to the sphere as soon as its impact
kinetic energy is less than a threshold which behaves linearly as a
function of the small particle size. The relevant parameter is thus
$\gamma = \min_n v_n^{\rm coll}/(U\,St^{1/4})$, where the minimum is
over all experienced collisions. For given settings (type of particles
and of fluid, nature of the sphere surface) accretion occurs when
$\gamma$ is less than a fixed dimensionless critical value
$\gamma_\star$.
\begin{figure}[h]
  \includegraphics[width=\columnwidth]{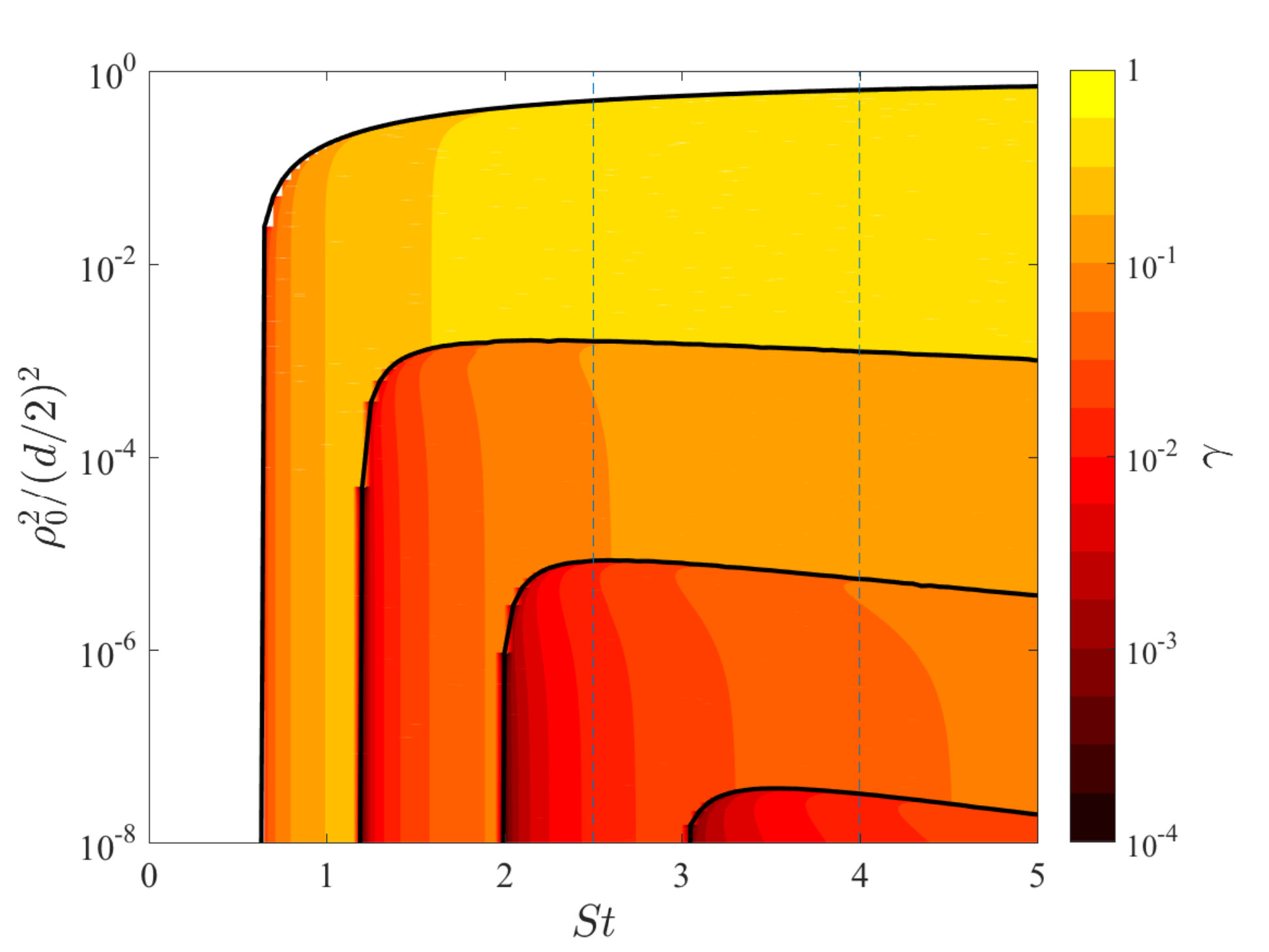}
  \vspace{-20pt}
  \caption{Minimal value of the impact parameter
    $\gamma = \min_n v_n^{\rm coll}/(U\,St^{1/4})$ (colored background
    and level lines) as a function of the particles Stokes number $St$
    and of the square $\rho_0^2$ of their initial distance to the
    symmetry axis in the Stokes creeping flow and for a restitution
    coefficient $e=1$.  The two vertical lines correspond to the cut
    shown in Fig.~\ref{fig:vcoll_fn_rho0_multicoll}.}
  \label{fig:efficiency_stokes_energy}
\end{figure}
Figure~\ref{fig:efficiency_stokes_energy} shows in the case of purely
elastic particles ($e=1$) how the impact parameter $\gamma$ depends
upon both the particles Stokes number and their initial distance
$\rho_0$ from the symmetry axis. We have represented here the minimal
value of $\gamma$ that a particle experiences when it has multiple
bounces, such as a part of the graph that is below a fixed value
$\gamma_\star$ defines particles which are actually sticking to the
sphere. The white area in the left/top of the figure corresponds to
values of $\rho_0$ and $St$ for which no collision occur and $\gamma$
is not defined. This area is delimited by the collision efficiency,
that is by the curve $\rho_0^2/(d/2)^2 = \mathcal{E}_{\rm coll}(St)$.
Below it, in the colored area, at least one collision occurs. This
area is itself divided into several islands (here four) associated to
successive rebounds of the particles. Each of these islands is bounded
from above and the left by a critical curve which defines the set of
parameters for which particles are indeed experiencing at least $n$
collisions.  The behavior of these curves as $\rho_0\to0$ defines a
critical value of the Stokes number $St_{\rm c}^{(n)}$ below which no
particle experiences $n$ collision. Of course, $St_{\rm c}^{(1)}$ is
given by the critical Stokes number $St_{\rm c}$ discussed in
Sec.~\ref{sec:coll-eff}. The typical values of $\gamma$ are decreasing
from one island to that below, because of the dissipation occuring
between successive bounces.  In each island, $\gamma$ is minimal close
to the upper/left boundary curve. This is because particles with such
settings are exactly falling on a critical trajectory: They approach
asymptotically the sphere and thus touch at an infinite time with a
vanishing velocity.  For the data shown in
Fig.~\ref{fig:efficiency_stokes_energy}, the impact parameter $\gamma$
decreases when $St$ increases. We however expect that at much larger
values of $St$, the impact velocity stabilizes to
$v^{\rm coll} \simeq U$, so that the impact parameter should decrease
as $\gamma\simeq St^{-1/4}$.

\begin{figure}[h]
  \includegraphics[width=\columnwidth]{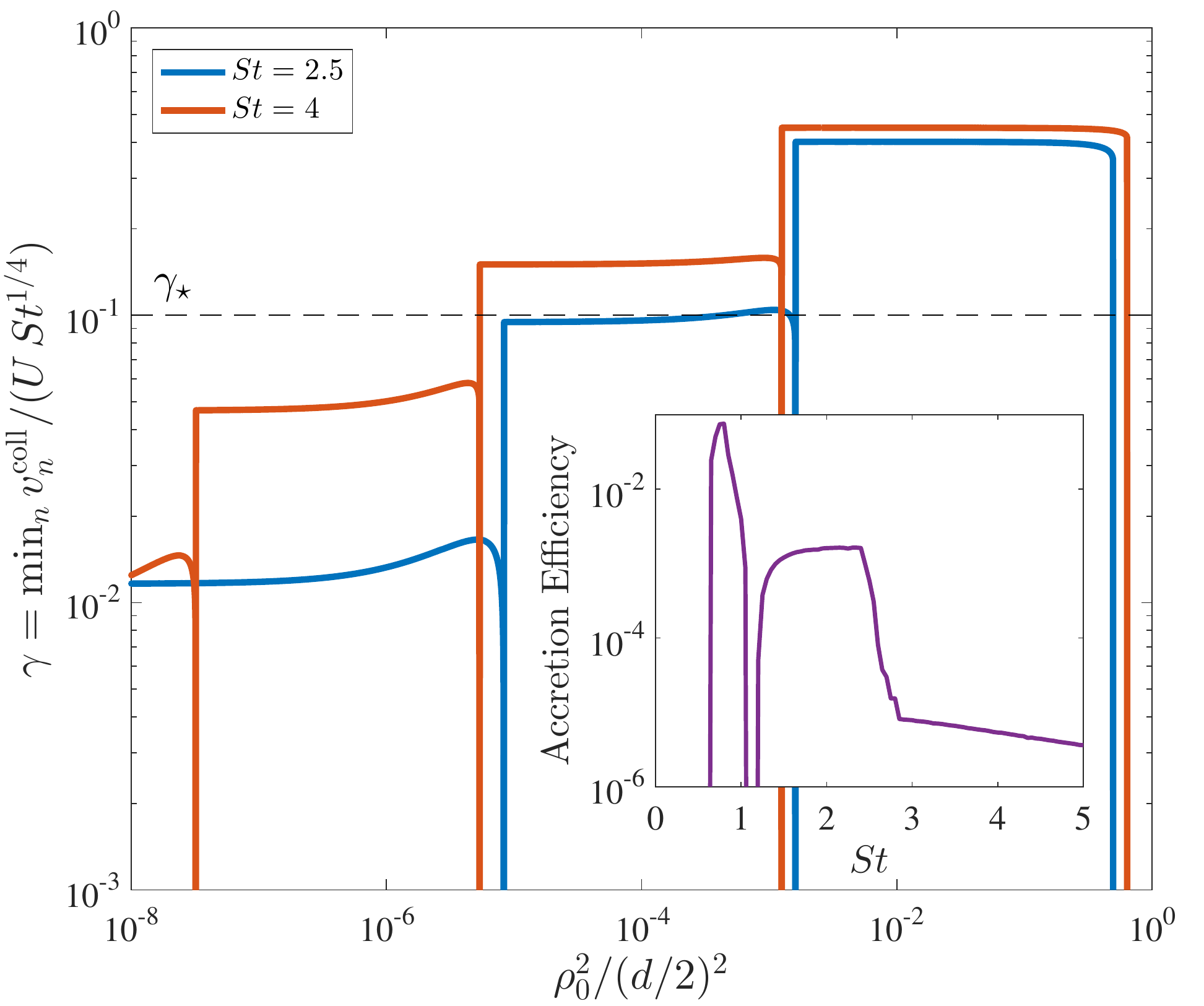}
  \vspace{-20pt}
  \caption{Impact parameter $\gamma$ for the Stokes flow and elastic
    particles ($e=1$) for two values of the Stokes number ($St=1.5$
    and $St=4$). The horizontal dashed line correspond to a specific
    choice $\gamma_\star= 0.1$ of the accretion critical
    parameter. All particles such that $\gamma<\gamma_\star$ stick to
    the sphere. The inset represents for the same settings the
    accretion efficiency associated to this specific value of
    $\gamma_\star$, as a function of the Stokes number.}
  \label{fig:vcoll_fn_rho0_multicoll}
\end{figure}
To get a better grasp on the interpretation of
Fig.~\ref{fig:efficiency_stokes_energy} in terms of accretion
efficiency, we have represented in
Fig.~\ref{fig:vcoll_fn_rho0_multicoll} two cuts for $St=2.5$ and
$St = 4$ of the impact parameter $\gamma$ as a function of
$\rho_0^2$. The rebounds islands now appear as steps.  Assume that we
fix the value $\gamma_\star$ of the critical impact parameter. The
accreting particles are those which were located at an initial
distance $\rho_0$ from the symmetric axis satisfying
$\gamma(\rho) < \gamma_\star$. The accretion efficiency is then
obtained by integrating the infinitesimal flux
$\propto \rho_0\, \mathrm{d}\rho_0 \propto \mathrm{d}\rho_0^2$
associated to these values. As an illustration, we consider
$\gamma_\star = 0.1$ (horizontal dashed line in
Fig.~\ref{fig:vcoll_fn_rho0_multicoll}). For $St=4$, all particles
colliding three times or more stick to the sphere. In addition, a
small fraction of the particles colliding once or twice will stick
when they are close to the critical trajectories. This contribution
corresponds to the steep dips separating successive plateaux. In the
case $St=2.5$, again all particles colliding at least three times will
accrete, together with those corresponding to the neighborhood of
critical trajectories. However an additional fraction of the particles
colliding only twice will stick to the sphere.  Putting together all
these contribution leads to an evaluation of the accretion rate, that
is of the ratio between the particles actually sticking to the sphere
and those included upstream in the sphere cross-section. This
efficiency is shown as an inset in
Fig.~\ref{fig:vcoll_fn_rho0_multicoll}, again for Stokes' flow, $e=1$,
and the specific choice $\gamma_\star=0.1$. The resulting curve has a
non-trivial dependence upon $St$ resulting from the various possible
behaviors described above.

\begin{figure}[h]
  \includegraphics[width=\columnwidth]{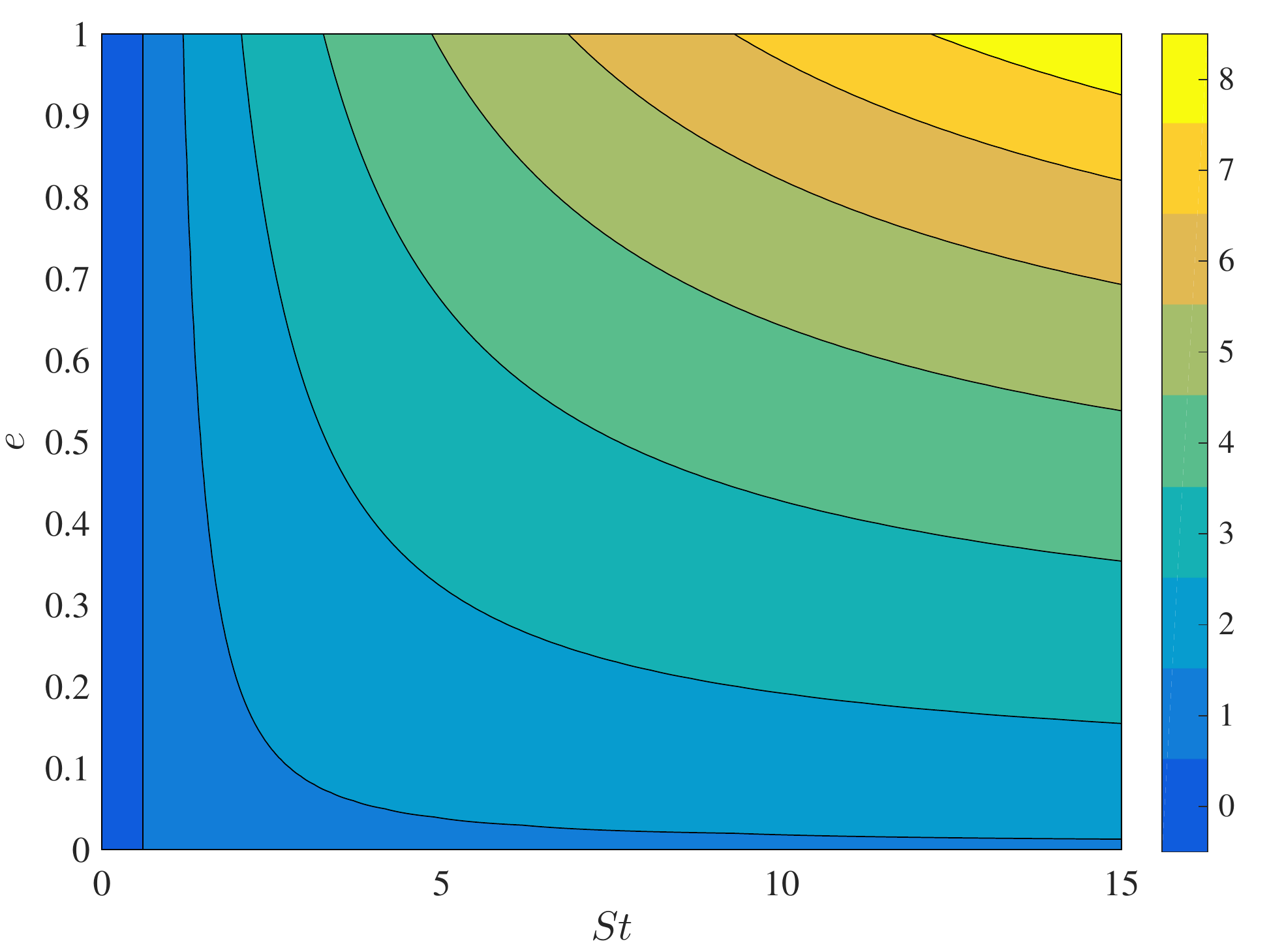}
  \vspace{-20pt}
  \caption{Maximum number of collisions experienced by particles in
    the Stokes flow, as a function of the Stokes number $St$ and of
    the restitution coefficient $e$. The black lines show the
    dependence upon $e$ of the critical Stokes number $St_c^{(n)}$
    above which particles experience at least $n$ bounces.}
  \label{fig:max_nb_collision_stokesflow}
\end{figure}

Notice that we have here chosen to represent data for the elastic case
$e=1$.  Lesser values of the restitution leads to qualitatively
equivalent results with of course a stronger depletion of impact
velocities from one bounce to the next but simultaneously, lesser
particles experiencing such a rebound. Because of this competition,
inelasticity can either enhance or reduce adhesion and its effect is
non monotonic. The dependence upon $e$ of the number of collisions
that can be experienced by given particles is illustrated in
Fig.~\ref{fig:max_nb_collision_stokesflow} for Stokes' creeping
flow. The black lines separating the various colored areas are given
by the value of the $n$-collisions critical Stokes numbers
$St_{\rm c}^{(n)}$. One indeed observes that for a given value of the
Stokes number, the maximal number of successive rebounds decreases as
a function of $e$.

\section{Concluding remarks}

We have here focused on the accretion of small particles by a large
fixed sphere embedded in a mean flow and we have investigated the
effect of inelastic collisions with the sphere. We have characterized
the collision efficiency as a function of the two dimensionless
parameters (Reynolds and Stokes numbers).  In particular, there is a
critical Stokes number below which no collision occurs and we have
observed that the collision efficiency depends on the Reynolds number
based on the sphere diameter only through the value of this critical
Stokes number. Besides, our theoretical analysis has demonstrated that
successive inelastic collisions do not lead to any kind of inelastic
collapse since multiple bounces do not provide enough energy
dissipation for the particles to stick to the surface within a finite
time. Yet, adding a refined criteria which accounts for
particle-surface microphysical interactions can change significantly
these results. Here, we have chosen to assume that the particles can
stick if their kinetic energy at impact is below a certain threshold.
Numerical simulations have shown that successive bounces can possibly
enhance accretion regardless of the Reynolds number.

The present numerical simulations also provide interesting qualitative
results.  It has indeed been shown that, in the case of a Stokes flow,
the accretion efficiency has a non-trivial behavior as a function of
the Stokes number for a given value of the critical impact parameter
($\gamma_\star$ below which particles stick to the surface).  In fact,
it appears that there is a 'selective' (or preferential) accretion of
specific particle sizes: The accretion efficiency is relatively high
at low Stokes number but drops significantly for a narrow range of
particle Stokes numbers (around $St=1$).  In the context of wet
deposition, such a selective accretion can prove to be dramatic in the
scavenging of ultrafine particles (within the micrometer range) which
are the most dangerous to health. For that reason, it will be also
interesting to investigate how the accretion efficiency evolves as a
function of the particle Stokes number for other flows (Euler flow,
turbulent flows at various Reynolds number) and to verify whether such
a selective accretion remains valid in such cases.

These results raise additional questions on the accretion of small
particles by a large sphere, among which:
\begin{enumerate}
\item[i.] What is the effect of diffusion? It has indeed been shown
  that multiple bounces cannot provide any kind of inelastic collapse
  in the absence of molecular diffusion. In principle, adding
  diffusion will give two effects: On the one hand it will alter the
  results at small values of the Stokes number since the existence of
  a critical Stokes number disappears. On the other hand, diffusion
  will grant inelastic collapse for sufficiently small values of the
  restitution coefficient.
\item[ii.] What is the effect of more realistic particle-sphere
  interactions?  The present simulations have been performed in the
  simple case of a flow past a sphere without accounting for specific
  particle-surface interactions such as: electro-magnetic forces or
  gravitation (in the context of astrophysical flows), surface tension
  or capillary forces (in the context of bubble/droplet growth).
  Besides, more realistic outcomes of collisions can also be
  investigated by taking into account friction forces and their effect
  on particle rotation.
\item[iii.] What is the effect of geometry? A large part of the
  present results, even at a qualitative level, have been obtained in
  the case of a flow past a sphere. In that case, it has been seen
  that bouncing particles can escape downstream if their tangential
  velocity after impact is high enough to prevent them from coming
  back to the sphere surface. This situation is specific to the case
  of a flow past a sphere but can be significantly different in the
  case of a different geometry such as a flat infinite boundary. This
  will be the topic of future studies.  In particular, we aim at
  analyzing the effect of successive bounces on the inelastic collapse
  in the case of particles impacting a flat boundary, such as a
  channel flow.
\end{enumerate}
  
\bibliography{biblio}

\begin{thebibliography}{42}%
\makeatletter
\providecommand \@ifxundefined [1]{%
 \@ifx{#1\undefined}
}%
\providecommand \@ifnum [1]{%
 \ifnum #1\expandafter \@firstoftwo
 \else \expandafter \@secondoftwo
 \fi
}%
\providecommand \@ifx [1]{%
 \ifx #1\expandafter \@firstoftwo
 \else \expandafter \@secondoftwo
 \fi
}%
\providecommand \natexlab [1]{#1}%
\providecommand \enquote  [1]{``#1''}%
\providecommand \bibnamefont  [1]{#1}%
\providecommand \bibfnamefont [1]{#1}%
\providecommand \citenamefont [1]{#1}%
\providecommand \href@noop [0]{\@secondoftwo}%
\providecommand \href [0]{\begingroup \@sanitize@url \@href}%
\providecommand \@href[1]{\@@startlink{#1}\@@href}%
\providecommand \@@href[1]{\endgroup#1\@@endlink}%
\providecommand \@sanitize@url [0]{\catcode `\\12\catcode `\$12\catcode
  `\&12\catcode `\#12\catcode `\^12\catcode `\_12\catcode `\%12\relax}%
\providecommand \@@startlink[1]{}%
\providecommand \@@endlink[0]{}%
\providecommand \url  [0]{\begingroup\@sanitize@url \@url }%
\providecommand \@url [1]{\endgroup\@href {#1}{\urlprefix }}%
\providecommand \urlprefix  [0]{URL }%
\providecommand \Eprint [0]{\href }%
\providecommand \doibase [0]{http://dx.doi.org/}%
\providecommand \selectlanguage [0]{\@gobble}%
\providecommand \bibinfo  [0]{\@secondoftwo}%
\providecommand \bibfield  [0]{\@secondoftwo}%
\providecommand \translation [1]{[#1]}%
\providecommand \BibitemOpen [0]{}%
\providecommand \bibitemStop [0]{}%
\providecommand \bibitemNoStop [0]{.\EOS\space}%
\providecommand \EOS [0]{\spacefactor3000\relax}%
\providecommand \BibitemShut  [1]{\csname bibitem#1\endcsname}%
\let\auto@bib@innerbib\@empty
\bibitem [{\citenamefont {Henry}\ \emph {et~al.}(2012)\citenamefont {Henry},
  \citenamefont {Minier},\ and\ \citenamefont
  {Lef{\`e}vre}}]{henry2012towards}%
  \BibitemOpen
  \bibfield  {author} {\bibinfo {author} {\bibfnamefont {C.}~\bibnamefont
  {Henry}}, \bibinfo {author} {\bibfnamefont {J.-P.}\ \bibnamefont {Minier}}, \
  and\ \bibinfo {author} {\bibfnamefont {G.}~\bibnamefont {Lef{\`e}vre}},\
  }\href {http://www.sciencedirect.com/science/article/pii/S0001868612001418}
  {\bibfield  {journal} {\bibinfo  {journal} {Adv. Colloid Interface Sci.}\
  }\textbf {\bibinfo {volume} {185}},\ \bibinfo {pages} {34} (\bibinfo {year}
  {2012})}\BibitemShut {NoStop}%
\bibitem [{\citenamefont {Pruppacher}\ and\ \citenamefont
  {Klett}(1997)}]{pruppacher-klett:1997}%
  \BibitemOpen
  \bibfield  {author} {\bibinfo {author} {\bibfnamefont {H.}~\bibnamefont
  {Pruppacher}}\ and\ \bibinfo {author} {\bibfnamefont {J.}~\bibnamefont
  {Klett}},\ }\href@noop {} {\emph {\bibinfo {title} {Microphysics of Clouds
  and Precipitation}}}\ (\bibinfo  {publisher} {Kluwer Academic},\ \bibinfo
  {address} {Dordrecht},\ \bibinfo {year} {1997})\BibitemShut {NoStop}%
\bibitem [{\citenamefont {Lissauer}(1993)}]{lissauer:1993}%
  \BibitemOpen
  \bibfield  {author} {\bibinfo {author} {\bibfnamefont {J.~J.}\ \bibnamefont
  {Lissauer}},\ }\href@noop {} {\bibfield  {journal} {\bibinfo  {journal}
  {Annu. Rev. Astron. Astrophys.}\ }\textbf {\bibinfo {volume} {31}},\ \bibinfo
  {pages} {129} (\bibinfo {year} {1993})}\BibitemShut {NoStop}%
\bibitem [{\citenamefont {Bec}\ \emph {et~al.}(2013)\citenamefont {Bec},
  \citenamefont {Musacchio},\ and\ \citenamefont {Ray}}]{bec2013sticky}%
  \BibitemOpen
  \bibfield  {author} {\bibinfo {author} {\bibfnamefont {J.}~\bibnamefont
  {Bec}}, \bibinfo {author} {\bibfnamefont {S.}~\bibnamefont {Musacchio}}, \
  and\ \bibinfo {author} {\bibfnamefont {S.~S.}\ \bibnamefont {Ray}},\ }\href
  {https://journals.aps.org/pre/abstract/10.1103/PhysRevE.87.063013} {\bibfield
   {journal} {\bibinfo  {journal} {Phys. Rev. E}\ }\textbf {\bibinfo {volume}
  {87}},\ \bibinfo {pages} {063013} (\bibinfo {year} {2013})}\BibitemShut
  {NoStop}%
\bibitem [{\citenamefont {Belan}\ \emph {et~al.}(2014)\citenamefont {Belan},
  \citenamefont {Fouxon},\ and\ \citenamefont
  {Falkovich}}]{belan2014localization}%
  \BibitemOpen
  \bibfield  {author} {\bibinfo {author} {\bibfnamefont {S.}~\bibnamefont
  {Belan}}, \bibinfo {author} {\bibfnamefont {I.}~\bibnamefont {Fouxon}}, \
  and\ \bibinfo {author} {\bibfnamefont {G.}~\bibnamefont {Falkovich}},\ }\href
  {https://journals.aps.org/prl/abstract/10.1103/PhysRevLett.112.234502}
  {\bibfield  {journal} {\bibinfo  {journal} {Phys. Rev. Lett.}\ }\textbf
  {\bibinfo {volume} {112}},\ \bibinfo {pages} {234502} (\bibinfo {year}
  {2014})}\BibitemShut {NoStop}%
\bibitem [{\citenamefont {Belan}(2016)}]{belan2016concentration}%
  \BibitemOpen
  \bibfield  {author} {\bibinfo {author} {\bibfnamefont {S.}~\bibnamefont
  {Belan}},\ }\href
  {http://www.sciencedirect.com/science/article/pii/S0378437115008043}
  {\bibfield  {journal} {\bibinfo  {journal} {Physica A}\ }\textbf {\bibinfo
  {volume} {443}},\ \bibinfo {pages} {128} (\bibinfo {year}
  {2016})}\BibitemShut {NoStop}%
\bibitem [{\citenamefont {Belan}\ \emph {et~al.}(2016)\citenamefont {Belan},
  \citenamefont {Chernykh}, \citenamefont {Lebedev},\ and\ \citenamefont
  {Falkovich}}]{belan2016inelastic}%
  \BibitemOpen
  \bibfield  {author} {\bibinfo {author} {\bibfnamefont {S.}~\bibnamefont
  {Belan}}, \bibinfo {author} {\bibfnamefont {A.}~\bibnamefont {Chernykh}},
  \bibinfo {author} {\bibfnamefont {V.}~\bibnamefont {Lebedev}}, \ and\
  \bibinfo {author} {\bibfnamefont {G.}~\bibnamefont {Falkovich}},\ }\href
  {https://journals.aps.org/pre/abstract/10.1103/PhysRevE.93.052206} {\bibfield
   {journal} {\bibinfo  {journal} {Phys. Rev. E}\ }\textbf {\bibinfo {volume}
  {93}},\ \bibinfo {pages} {052206} (\bibinfo {year} {2016})}\BibitemShut
  {NoStop}%
\bibitem [{\citenamefont {Cornell}\ \emph {et~al.}(1998)\citenamefont
  {Cornell}, \citenamefont {Swift},\ and\ \citenamefont
  {Bray}}]{cornell1998inelastic}%
  \BibitemOpen
  \bibfield  {author} {\bibinfo {author} {\bibfnamefont {S.~J.}\ \bibnamefont
  {Cornell}}, \bibinfo {author} {\bibfnamefont {M.~R.}\ \bibnamefont {Swift}},
  \ and\ \bibinfo {author} {\bibfnamefont {A.~J.}\ \bibnamefont {Bray}},\
  }\href {https://journals.aps.org/prl/abstract/10.1103/PhysRevLett.81.1142}
  {\bibfield  {journal} {\bibinfo  {journal} {Phys. Rev. Lett.}\ }\textbf
  {\bibinfo {volume} {81}},\ \bibinfo {pages} {1142} (\bibinfo {year}
  {1998})}\BibitemShut {NoStop}%
\bibitem [{\citenamefont {Cecconi}\ \emph {et~al.}(2003)\citenamefont
  {Cecconi}, \citenamefont {Puglisi}, \citenamefont {Marconi},\ and\
  \citenamefont {Vulpiani}}]{cecconi2003noise}%
  \BibitemOpen
  \bibfield  {author} {\bibinfo {author} {\bibfnamefont {F.}~\bibnamefont
  {Cecconi}}, \bibinfo {author} {\bibfnamefont {A.}~\bibnamefont {Puglisi}},
  \bibinfo {author} {\bibfnamefont {U.~M.~B.}\ \bibnamefont {Marconi}}, \ and\
  \bibinfo {author} {\bibfnamefont {A.}~\bibnamefont {Vulpiani}},\ }\href
  {https://journals.aps.org/prl/abstract/10.1103/PhysRevLett.90.064301}
  {\bibfield  {journal} {\bibinfo  {journal} {Phys. Rev. Lett.}\ }\textbf
  {\bibinfo {volume} {90}},\ \bibinfo {pages} {064301} (\bibinfo {year}
  {2003})}\BibitemShut {NoStop}%
\bibitem [{\citenamefont {Mircea}\ \emph {et~al.}(2000)\citenamefont {Mircea},
  \citenamefont {Stefan},\ and\ \citenamefont
  {Fuzzi}}]{mircea2000precipitation}%
  \BibitemOpen
  \bibfield  {author} {\bibinfo {author} {\bibfnamefont {M.}~\bibnamefont
  {Mircea}}, \bibinfo {author} {\bibfnamefont {S.}~\bibnamefont {Stefan}}, \
  and\ \bibinfo {author} {\bibfnamefont {S.}~\bibnamefont {Fuzzi}},\ }\href
  {http://www.sciencedirect.com/science/article/pii/S1352231000001990}
  {\bibfield  {journal} {\bibinfo  {journal} {Atmos. Environ.}\ }\textbf
  {\bibinfo {volume} {34}},\ \bibinfo {pages} {5169} (\bibinfo {year}
  {2000})}\BibitemShut {NoStop}%
\bibitem [{\citenamefont {Berthet}\ \emph {et~al.}(2010)\citenamefont
  {Berthet}, \citenamefont {Leriche}, \citenamefont {Pinty}, \citenamefont
  {Cuesta},\ and\ \citenamefont {Pigeon}}]{berthet2010scavenging}%
  \BibitemOpen
  \bibfield  {author} {\bibinfo {author} {\bibfnamefont {S.}~\bibnamefont
  {Berthet}}, \bibinfo {author} {\bibfnamefont {M.}~\bibnamefont {Leriche}},
  \bibinfo {author} {\bibfnamefont {J.-P.}\ \bibnamefont {Pinty}}, \bibinfo
  {author} {\bibfnamefont {J.}~\bibnamefont {Cuesta}}, \ and\ \bibinfo {author}
  {\bibfnamefont {G.}~\bibnamefont {Pigeon}},\ }\href
  {http://www.sciencedirect.com/science/article/pii/S0169809509002725}
  {\bibfield  {journal} {\bibinfo  {journal} {Atmos. Res.}\ }\textbf {\bibinfo
  {volume} {96}},\ \bibinfo {pages} {325} (\bibinfo {year} {2010})}\BibitemShut
  {NoStop}%
\bibitem [{\citenamefont {Levy}\ \emph {et~al.}(2013)\citenamefont {Levy},
  \citenamefont {Horowitz}, \citenamefont {Schwarzkopf}, \citenamefont {Ming},
  \citenamefont {Golaz}, \citenamefont {Naik},\ and\ \citenamefont
  {Ramaswamy}}]{levy2013roles}%
  \BibitemOpen
  \bibfield  {author} {\bibinfo {author} {\bibfnamefont {H.}~\bibnamefont
  {Levy}}, \bibinfo {author} {\bibfnamefont {L.~W.}\ \bibnamefont {Horowitz}},
  \bibinfo {author} {\bibfnamefont {M.~D.}\ \bibnamefont {Schwarzkopf}},
  \bibinfo {author} {\bibfnamefont {Y.}~\bibnamefont {Ming}}, \bibinfo {author}
  {\bibfnamefont {J.-C.}\ \bibnamefont {Golaz}}, \bibinfo {author}
  {\bibfnamefont {V.}~\bibnamefont {Naik}}, \ and\ \bibinfo {author}
  {\bibfnamefont {V.}~\bibnamefont {Ramaswamy}},\ }\href
  {http://onlinelibrary.wiley.com/doi/10.1002/jgrd.50192/full} {\bibfield
  {journal} {\bibinfo  {journal} {J. Geophys. Res.}\ }\textbf {\bibinfo
  {volume} {118}},\ \bibinfo {pages} {4521} (\bibinfo {year}
  {2013})}\BibitemShut {NoStop}%
\bibitem [{\citenamefont {Guo}\ \emph {et~al.}(2016)\citenamefont {Guo},
  \citenamefont {Zhang}, \citenamefont {Lin}, \citenamefont {Zeng},
  \citenamefont {Liu}, \citenamefont {Xiao}, \citenamefont {Rutherford},
  \citenamefont {You},\ and\ \citenamefont {Ma}}]{guo2016washout}%
  \BibitemOpen
  \bibfield  {author} {\bibinfo {author} {\bibfnamefont {L.-C.}\ \bibnamefont
  {Guo}}, \bibinfo {author} {\bibfnamefont {Y.}~\bibnamefont {Zhang}}, \bibinfo
  {author} {\bibfnamefont {H.}~\bibnamefont {Lin}}, \bibinfo {author}
  {\bibfnamefont {W.}~\bibnamefont {Zeng}}, \bibinfo {author} {\bibfnamefont
  {T.}~\bibnamefont {Liu}}, \bibinfo {author} {\bibfnamefont {J.}~\bibnamefont
  {Xiao}}, \bibinfo {author} {\bibfnamefont {S.}~\bibnamefont {Rutherford}},
  \bibinfo {author} {\bibfnamefont {J.}~\bibnamefont {You}}, \ and\ \bibinfo
  {author} {\bibfnamefont {W.}~\bibnamefont {Ma}},\ }\href
  {http://www.sciencedirect.com/science/article/pii/S0269749116303803}
  {\bibfield  {journal} {\bibinfo  {journal} {Environ. Pollut.}\ }\textbf
  {\bibinfo {volume} {215}},\ \bibinfo {pages} {195} (\bibinfo {year}
  {2016})}\BibitemShut {NoStop}%
\bibitem [{\citenamefont {Beard}(1974)}]{beard1974experimental}%
  \BibitemOpen
  \bibfield  {author} {\bibinfo {author} {\bibfnamefont {K.}~\bibnamefont
  {Beard}},\ }\href
  {http://journals.ametsoc.org/doi/abs/10.1175/1520-0469(1974)031%3C1595:EANCEF%3E2.0.CO;2}
  {\bibfield  {journal} {\bibinfo  {journal} {J.\ Atmos. Sci.}\ }\textbf
  {\bibinfo {volume} {31}},\ \bibinfo {pages} {1595} (\bibinfo {year}
  {1974})}\BibitemShut {NoStop}%
\bibitem [{\citenamefont {Beard}\ and\ \citenamefont
  {Grover}(1974)}]{beard1974numerical}%
  \BibitemOpen
  \bibfield  {author} {\bibinfo {author} {\bibfnamefont {K.}~\bibnamefont
  {Beard}}\ and\ \bibinfo {author} {\bibfnamefont {S.}~\bibnamefont {Grover}},\
  }\href
  {http://journals.ametsoc.org/doi/abs/10.1175/1520-0469(1974)031%3C0543:NCEFSR%3E2.0.CO%3B2}
  {\bibfield  {journal} {\bibinfo  {journal} {J. Atmos. Sci.}\ }\textbf
  {\bibinfo {volume} {31}},\ \bibinfo {pages} {543} (\bibinfo {year}
  {1974})}\BibitemShut {NoStop}%
\bibitem [{\citenamefont {Ardon-Dryer}\ \emph {et~al.}(2015)\citenamefont
  {Ardon-Dryer}, \citenamefont {Huang},\ and\ \citenamefont
  {Cziczo}}]{ardon2015laboratory}%
  \BibitemOpen
  \bibfield  {author} {\bibinfo {author} {\bibfnamefont {K.}~\bibnamefont
  {Ardon-Dryer}}, \bibinfo {author} {\bibfnamefont {Y.-W.}\ \bibnamefont
  {Huang}}, \ and\ \bibinfo {author} {\bibfnamefont {D.~J.}\ \bibnamefont
  {Cziczo}},\ }\href
  {http://www.atmos-chem-phys.net/15/9159/2015/acp-15-9159-2015.pdf} {\bibfield
   {journal} {\bibinfo  {journal} {Atmos. Chem. Phys.}\ }\textbf {\bibinfo
  {volume} {15}},\ \bibinfo {pages} {9159} (\bibinfo {year}
  {2015})}\BibitemShut {NoStop}%
\bibitem [{\citenamefont {Lemaitre}\ \emph {et~al.}(2017)\citenamefont
  {Lemaitre}, \citenamefont {Querel}, \citenamefont {Monier}, \citenamefont
  {Menard}, \citenamefont {Porcheron},\ and\ \citenamefont
  {Flossmann}}]{lemaitre2017experimental}%
  \BibitemOpen
  \bibfield  {author} {\bibinfo {author} {\bibfnamefont {P.}~\bibnamefont
  {Lemaitre}}, \bibinfo {author} {\bibfnamefont {A.}~\bibnamefont {Querel}},
  \bibinfo {author} {\bibfnamefont {M.}~\bibnamefont {Monier}}, \bibinfo
  {author} {\bibfnamefont {T.}~\bibnamefont {Menard}}, \bibinfo {author}
  {\bibfnamefont {E.}~\bibnamefont {Porcheron}}, \ and\ \bibinfo {author}
  {\bibfnamefont {A.~I.}\ \bibnamefont {Flossmann}},\ }\href
  {http://www.atmos-chem-phys.net/17/4159/2017/} {\bibfield  {journal}
  {\bibinfo  {journal} {Atmos. Chem. Phys.}\ }\textbf {\bibinfo {volume} {17}}
  (\bibinfo {year} {2017})}\BibitemShut {NoStop}%
\bibitem [{\citenamefont {Chen}\ \emph {et~al.}(2016)\citenamefont {Chen},
  \citenamefont {Hu}, \citenamefont {Liu}, \citenamefont {Xu}, \citenamefont
  {Yang}, \citenamefont {Xu},\ and\ \citenamefont {Chen}}]{chen2016beyond}%
  \BibitemOpen
  \bibfield  {author} {\bibinfo {author} {\bibfnamefont {R.}~\bibnamefont
  {Chen}}, \bibinfo {author} {\bibfnamefont {B.}~\bibnamefont {Hu}}, \bibinfo
  {author} {\bibfnamefont {Y.}~\bibnamefont {Liu}}, \bibinfo {author}
  {\bibfnamefont {J.}~\bibnamefont {Xu}}, \bibinfo {author} {\bibfnamefont
  {G.}~\bibnamefont {Yang}}, \bibinfo {author} {\bibfnamefont {D.}~\bibnamefont
  {Xu}}, \ and\ \bibinfo {author} {\bibfnamefont {C.}~\bibnamefont {Chen}},\
  }\href {http://www.sciencedirect.com/science/article/pii/S0304416516300745}
  {\bibfield  {journal} {\bibinfo  {journal} {Biochim. Biophys. Acta}\ }\textbf
  {\bibinfo {volume} {1860}},\ \bibinfo {pages} {2844} (\bibinfo {year}
  {2016})}\BibitemShut {NoStop}%
\bibitem [{\citenamefont {Phillips}\ and\ \citenamefont
  {Kaye}(1999)}]{phillips1999influence}%
  \BibitemOpen
  \bibfield  {author} {\bibinfo {author} {\bibfnamefont {C.}~\bibnamefont
  {Phillips}}\ and\ \bibinfo {author} {\bibfnamefont {S.}~\bibnamefont
  {Kaye}},\ }\href
  {http://www.sciencedirect.com/science/article/pii/S0021850298007666}
  {\bibfield  {journal} {\bibinfo  {journal} {J. Aerosol Sci.}\ }\textbf
  {\bibinfo {volume} {30}},\ \bibinfo {pages} {709} (\bibinfo {year}
  {1999})}\BibitemShut {NoStop}%
\bibitem [{\citenamefont {Slinn}(1983)}]{slinn1983precipitation}%
  \BibitemOpen
  \bibfield  {author} {\bibinfo {author} {\bibfnamefont {W.}~\bibnamefont
  {Slinn}},\ }\href@noop {} {\bibfield  {journal} {\bibinfo  {journal}
  {Atmospheric Sciences and Power Production}\ }\textbf {\bibinfo {volume}
  {Chap.11}} (\bibinfo {year} {1983})}\BibitemShut {NoStop}%
\bibitem [{\citenamefont {Sellentin}\ \emph {et~al.}(2013)\citenamefont
  {Sellentin}, \citenamefont {Ramsey}, \citenamefont {Windmark},\ and\
  \citenamefont {Dullemond}}]{sellentin2013quantification}%
  \BibitemOpen
  \bibfield  {author} {\bibinfo {author} {\bibfnamefont {E.}~\bibnamefont
  {Sellentin}}, \bibinfo {author} {\bibfnamefont {J.}~\bibnamefont {Ramsey}},
  \bibinfo {author} {\bibfnamefont {F.}~\bibnamefont {Windmark}}, \ and\
  \bibinfo {author} {\bibfnamefont {C.}~\bibnamefont {Dullemond}},\ }\href
  {http://www.aanda.org/articles/aa/full_html/2013/12/aa21587-13/aa21587-13.html}
  {\bibfield  {journal} {\bibinfo  {journal} {Astron. Astrophys.}\ }\textbf
  {\bibinfo {volume} {560}},\ \bibinfo {pages} {A96} (\bibinfo {year}
  {2013})}\BibitemShut {NoStop}%
\bibitem [{\citenamefont {Homann}\ \emph {et~al.}(2016)\citenamefont {Homann},
  \citenamefont {Guillot}, \citenamefont {Bec}, \citenamefont {Ormel},
  \citenamefont {Ida},\ and\ \citenamefont {Tanga}}]{homann2016effect}%
  \BibitemOpen
  \bibfield  {author} {\bibinfo {author} {\bibfnamefont {H.}~\bibnamefont
  {Homann}}, \bibinfo {author} {\bibfnamefont {T.}~\bibnamefont {Guillot}},
  \bibinfo {author} {\bibfnamefont {J.}~\bibnamefont {Bec}}, \bibinfo {author}
  {\bibfnamefont {C.}~\bibnamefont {Ormel}}, \bibinfo {author} {\bibfnamefont
  {S.}~\bibnamefont {Ida}}, \ and\ \bibinfo {author} {\bibfnamefont
  {P.}~\bibnamefont {Tanga}},\ }\href
  {http://www.aanda.org/articles/aa/full_html/2016/05/aa27344-15/aa27344-15.html}
  {\bibfield  {journal} {\bibinfo  {journal} {Astron. Astrophys.}\ }\textbf
  {\bibinfo {volume} {589}},\ \bibinfo {pages} {A129} (\bibinfo {year}
  {2016})}\BibitemShut {NoStop}%
\bibitem [{\citenamefont {Chokshi}\ \emph {et~al.}(1993)\citenamefont
  {Chokshi}, \citenamefont {Tielens},\ and\ \citenamefont
  {Hollenbach}}]{chokshi1993dust}%
  \BibitemOpen
  \bibfield  {author} {\bibinfo {author} {\bibfnamefont {A.}~\bibnamefont
  {Chokshi}}, \bibinfo {author} {\bibfnamefont {A.}~\bibnamefont {Tielens}}, \
  and\ \bibinfo {author} {\bibfnamefont {D.}~\bibnamefont {Hollenbach}},\
  }\href {http://adsabs.harvard.edu/full/1993ApJ...407..806C} {\bibfield
  {journal} {\bibinfo  {journal} {Astrophys. J.}\ }\textbf {\bibinfo {volume}
  {407}},\ \bibinfo {pages} {806} (\bibinfo {year} {1993})}\BibitemShut
  {NoStop}%
\bibitem [{\citenamefont {Windmark}\ \emph {et~al.}(2012)\citenamefont
  {Windmark}, \citenamefont {Birnstiel}, \citenamefont {Ormel},\ and\
  \citenamefont {Dullemond}}]{windmark2012breaking}%
  \BibitemOpen
  \bibfield  {author} {\bibinfo {author} {\bibfnamefont {F.}~\bibnamefont
  {Windmark}}, \bibinfo {author} {\bibfnamefont {T.}~\bibnamefont {Birnstiel}},
  \bibinfo {author} {\bibfnamefont {C.}~\bibnamefont {Ormel}}, \ and\ \bibinfo
  {author} {\bibfnamefont {C.~P.}\ \bibnamefont {Dullemond}},\ }\href
  {http://www.aanda.org/articles/aa/full_html/2012/08/aa20004-12/aa20004-12.html}
  {\bibfield  {journal} {\bibinfo  {journal} {Astron. Astrophys.}\ }\textbf
  {\bibinfo {volume} {544}},\ \bibinfo {pages} {L16} (\bibinfo {year}
  {2012})}\BibitemShut {NoStop}%
\bibitem [{\citenamefont {Garaud}\ \emph {et~al.}(2013)\citenamefont {Garaud},
  \citenamefont {Meru}, \citenamefont {Galvagni},\ and\ \citenamefont
  {Olczak}}]{garaud2013dust}%
  \BibitemOpen
  \bibfield  {author} {\bibinfo {author} {\bibfnamefont {P.}~\bibnamefont
  {Garaud}}, \bibinfo {author} {\bibfnamefont {F.}~\bibnamefont {Meru}},
  \bibinfo {author} {\bibfnamefont {M.}~\bibnamefont {Galvagni}}, \ and\
  \bibinfo {author} {\bibfnamefont {C.}~\bibnamefont {Olczak}},\ }\href
  {http://iopscience.iop.org/article/10.1088/0004-637X/764/2/146/meta}
  {\bibfield  {journal} {\bibinfo  {journal} {Astrophys. J.}\ }\textbf
  {\bibinfo {volume} {764}},\ \bibinfo {pages} {146} (\bibinfo {year}
  {2013})}\BibitemShut {NoStop}%
\bibitem [{\citenamefont {Mitra}\ \emph {et~al.}(2013)\citenamefont {Mitra},
  \citenamefont {Wettlaufer},\ and\ \citenamefont
  {Brandenburg}}]{mitra2013can}%
  \BibitemOpen
  \bibfield  {author} {\bibinfo {author} {\bibfnamefont {D.}~\bibnamefont
  {Mitra}}, \bibinfo {author} {\bibfnamefont {J.~S.}\ \bibnamefont
  {Wettlaufer}}, \ and\ \bibinfo {author} {\bibfnamefont {A.}~\bibnamefont
  {Brandenburg}},\ }\href
  {http://iopscience.iop.org/article/10.1088/0004-637X/773/2/120/meta}
  {\bibfield  {journal} {\bibinfo  {journal} {Astrophys. J.}\ }\textbf
  {\bibinfo {volume} {773}},\ \bibinfo {pages} {120} (\bibinfo {year}
  {2013})}\BibitemShut {NoStop}%
\bibitem [{\citenamefont {Hachem}\ \emph {et~al.}(2010)\citenamefont {Hachem},
  \citenamefont {Rivaux}, \citenamefont {Kloczko}, \citenamefont {Digonnet},\
  and\ \citenamefont {Coupez}}]{hachem2010stabilized}%
  \BibitemOpen
  \bibfield  {author} {\bibinfo {author} {\bibfnamefont {E.}~\bibnamefont
  {Hachem}}, \bibinfo {author} {\bibfnamefont {B.}~\bibnamefont {Rivaux}},
  \bibinfo {author} {\bibfnamefont {T.}~\bibnamefont {Kloczko}}, \bibinfo
  {author} {\bibfnamefont {H.}~\bibnamefont {Digonnet}}, \ and\ \bibinfo
  {author} {\bibfnamefont {T.}~\bibnamefont {Coupez}},\ }\href
  {http://www.sciencedirect.com/science/article/pii/S0021999110004237}
  {\bibfield  {journal} {\bibinfo  {journal} {J. Comput. Phys.}\ }\textbf
  {\bibinfo {volume} {229}},\ \bibinfo {pages} {8643} (\bibinfo {year}
  {2010})}\BibitemShut {NoStop}%
\bibitem [{\citenamefont {Drazin}(2002)}]{drazin2002introduction}%
  \BibitemOpen
  \bibfield  {author} {\bibinfo {author} {\bibfnamefont {P.}~\bibnamefont
  {Drazin}},\ }\href@noop {} {\emph {\bibinfo {title} {Introduction to
  Hydrodynamic Stability}}}\ (\bibinfo  {publisher} {Cambridge University
  Press},\ \bibinfo {address} {Cambridge},\ \bibinfo {year} {2002})\BibitemShut
  {NoStop}%
\bibitem [{\citenamefont {Fabre}\ \emph {et~al.}(2008)\citenamefont {Fabre},
  \citenamefont {Auguste},\ and\ \citenamefont
  {Magnaudet}}]{fabre2008bifurcations}%
  \BibitemOpen
  \bibfield  {author} {\bibinfo {author} {\bibfnamefont {D.}~\bibnamefont
  {Fabre}}, \bibinfo {author} {\bibfnamefont {F.}~\bibnamefont {Auguste}}, \
  and\ \bibinfo {author} {\bibfnamefont {J.}~\bibnamefont {Magnaudet}},\ }\href
  {http://aip.scitation.org/doi/abs/10.1063/1.2909609} {\bibfield  {journal}
  {\bibinfo  {journal} {Phys. Fluids}\ }\textbf {\bibinfo {volume} {20}},\
  \bibinfo {pages} {051702} (\bibinfo {year} {2008})}\BibitemShut {NoStop}%
\bibitem [{\citenamefont {Falkovich}(2011)}]{falkovich2011fluid}%
  \BibitemOpen
  \bibfield  {author} {\bibinfo {author} {\bibfnamefont {G.}~\bibnamefont
  {Falkovich}},\ }\href@noop {} {\emph {\bibinfo {title} {Fluid mechanics: {A}
  short course for physicists}}}\ (\bibinfo  {publisher} {Cambridge University
  Press},\ \bibinfo {address} {Cambridge},\ \bibinfo {year} {2011})\BibitemShut
  {NoStop}%
\bibitem [{\citenamefont {Ashgriz}\ and\ \citenamefont
  {Poo}(1990)}]{ashgriz1990coalescence}%
  \BibitemOpen
  \bibfield  {author} {\bibinfo {author} {\bibfnamefont {N.}~\bibnamefont
  {Ashgriz}}\ and\ \bibinfo {author} {\bibfnamefont {J.}~\bibnamefont {Poo}},\
  }\href {http://journals.cambridge.org/article_S0022112090003536} {\bibfield
  {journal} {\bibinfo  {journal} {J. Fluid Mech.}\ }\textbf {\bibinfo {volume}
  {221}},\ \bibinfo {pages} {183} (\bibinfo {year} {1990})}\BibitemShut
  {NoStop}%
\bibitem [{\citenamefont {Qian}\ and\ \citenamefont
  {Law}(1997)}]{qian1997regimes}%
  \BibitemOpen
  \bibfield  {author} {\bibinfo {author} {\bibfnamefont {J.}~\bibnamefont
  {Qian}}\ and\ \bibinfo {author} {\bibfnamefont {C.}~\bibnamefont {Law}},\
  }\href {http://journals.cambridge.org/article_S0022112096003722} {\bibfield
  {journal} {\bibinfo  {journal} {J. Fluid Mech.}\ }\textbf {\bibinfo {volume}
  {331}},\ \bibinfo {pages} {59} (\bibinfo {year} {1997})}\BibitemShut
  {NoStop}%
\bibitem [{\citenamefont {Saffman}\ and\ \citenamefont
  {Turner}(1956)}]{saffman1956collision}%
  \BibitemOpen
  \bibfield  {author} {\bibinfo {author} {\bibfnamefont {P.}~\bibnamefont
  {Saffman}}\ and\ \bibinfo {author} {\bibfnamefont {J.}~\bibnamefont
  {Turner}},\ }\href {http://journals.cambridge.org/article_S0022112056000020}
  {\bibfield  {journal} {\bibinfo  {journal} {J. Fluid Mech.}\ }\textbf
  {\bibinfo {volume} {1}},\ \bibinfo {pages} {16} (\bibinfo {year}
  {1956})}\BibitemShut {NoStop}%
\bibitem [{\citenamefont {Sommerfeld}\ and\ \citenamefont
  {Kuschel}(2016)}]{sommerfeld2016modelling}%
  \BibitemOpen
  \bibfield  {author} {\bibinfo {author} {\bibfnamefont {M.}~\bibnamefont
  {Sommerfeld}}\ and\ \bibinfo {author} {\bibfnamefont {M.}~\bibnamefont
  {Kuschel}},\ }\href
  {https://link.springer.com/article/10.1007/s00348-016-2249-y} {\bibfield
  {journal} {\bibinfo  {journal} {Exp. Fluids}\ }\textbf {\bibinfo {volume}
  {57}},\ \bibinfo {pages} {187} (\bibinfo {year} {2016})}\BibitemShut
  {NoStop}%
\bibitem [{\citenamefont {Rabe}\ \emph {et~al.}(2010)\citenamefont {Rabe},
  \citenamefont {Malet},\ and\ \citenamefont
  {Feuillebois}}]{rabe2010experimental}%
  \BibitemOpen
  \bibfield  {author} {\bibinfo {author} {\bibfnamefont {C.}~\bibnamefont
  {Rabe}}, \bibinfo {author} {\bibfnamefont {J.}~\bibnamefont {Malet}}, \ and\
  \bibinfo {author} {\bibfnamefont {F.}~\bibnamefont {Feuillebois}},\ }\href
  {http://aip.scitation.org/doi/abs/10.1063/1.3392768} {\bibfield  {journal}
  {\bibinfo  {journal} {Phys. Fluids}\ }\textbf {\bibinfo {volume} {22}},\
  \bibinfo {pages} {047101} (\bibinfo {year} {2010})}\BibitemShut {NoStop}%
\bibitem [{\citenamefont {Eggers}\ and\ \citenamefont
  {Villermaux}(2008)}]{eggers2008physics}%
  \BibitemOpen
  \bibfield  {author} {\bibinfo {author} {\bibfnamefont {J.}~\bibnamefont
  {Eggers}}\ and\ \bibinfo {author} {\bibfnamefont {E.}~\bibnamefont
  {Villermaux}},\ }\href
  {http://iopscience.iop.org/article/10.1088/0034-4885/71/3/036601/meta}
  {\bibfield  {journal} {\bibinfo  {journal} {Rep. Prog. Phys.}\ }\textbf
  {\bibinfo {volume} {71}},\ \bibinfo {pages} {036601} (\bibinfo {year}
  {2008})}\BibitemShut {NoStop}%
\bibitem [{\citenamefont {Clift}\ \emph {et~al.}(1978)\citenamefont {Clift},
  \citenamefont {Grace},\ and\ \citenamefont {Weber}}]{clift2005bubbles}%
  \BibitemOpen
  \bibfield  {author} {\bibinfo {author} {\bibfnamefont {R.}~\bibnamefont
  {Clift}}, \bibinfo {author} {\bibfnamefont {J.~R.}\ \bibnamefont {Grace}}, \
  and\ \bibinfo {author} {\bibfnamefont {M.~E.}\ \bibnamefont {Weber}},\
  }\href@noop {} {\emph {\bibinfo {title} {Bubbles, drops, and particles}}}\
  (\bibinfo  {publisher} {Academic Press},\ \bibinfo {address} {New York},\
  \bibinfo {year} {1978})\BibitemShut {NoStop}%
\bibitem [{\citenamefont {Chate}\ \emph {et~al.}(2003)\citenamefont {Chate},
  \citenamefont {Rao}, \citenamefont {Naik}, \citenamefont {Momin},
  \citenamefont {Safai},\ and\ \citenamefont {Ali}}]{chate2003scavenging}%
  \BibitemOpen
  \bibfield  {author} {\bibinfo {author} {\bibfnamefont {D.}~\bibnamefont
  {Chate}}, \bibinfo {author} {\bibfnamefont {P.}~\bibnamefont {Rao}}, \bibinfo
  {author} {\bibfnamefont {M.}~\bibnamefont {Naik}}, \bibinfo {author}
  {\bibfnamefont {G.}~\bibnamefont {Momin}}, \bibinfo {author} {\bibfnamefont
  {P.}~\bibnamefont {Safai}}, \ and\ \bibinfo {author} {\bibfnamefont
  {K.}~\bibnamefont {Ali}},\ }\href
  {http://www.sciencedirect.com.insu.bib.cnrs.fr/science/article/pii/S1352231003001626}
  {\bibfield  {journal} {\bibinfo  {journal} {Atmos. Environ.}\ }\textbf
  {\bibinfo {volume} {37}},\ \bibinfo {pages} {2477} (\bibinfo {year}
  {2003})}\BibitemShut {NoStop}%
\bibitem [{\citenamefont {Elimelech}\ \emph {et~al.}(2013)\citenamefont
  {Elimelech}, \citenamefont {Gregory},\ and\ \citenamefont
  {Jia}}]{elimelech2013particle}%
  \BibitemOpen
  \bibfield  {author} {\bibinfo {author} {\bibfnamefont {M.}~\bibnamefont
  {Elimelech}}, \bibinfo {author} {\bibfnamefont {J.}~\bibnamefont {Gregory}},
  \ and\ \bibinfo {author} {\bibfnamefont {X.}~\bibnamefont {Jia}},\
  }\href@noop {} {\emph {\bibinfo {title} {Particle deposition and aggregation:
  measurement, modelling and simulation}}}\ (\bibinfo  {publisher}
  {Butterworth-Heinemann},\ \bibinfo {address} {Oxford},\ \bibinfo {year}
  {2013})\BibitemShut {NoStop}%
\bibitem [{\citenamefont {Maximova}\ and\ \citenamefont
  {Dahl}(2006)}]{maximova2006environmental}%
  \BibitemOpen
  \bibfield  {author} {\bibinfo {author} {\bibfnamefont {N.}~\bibnamefont
  {Maximova}}\ and\ \bibinfo {author} {\bibfnamefont {O.}~\bibnamefont
  {Dahl}},\ }\href
  {http://www.sciencedirect.com/science/article/pii/S1359029406000380}
  {\bibfield  {journal} {\bibinfo  {journal} {Current opinion in colloid \&
  interface science}\ }\textbf {\bibinfo {volume} {11}},\ \bibinfo {pages}
  {246} (\bibinfo {year} {2006})}\BibitemShut {NoStop}%
\bibitem [{\citenamefont {Stronge}(2004)}]{stronge2004impact}%
  \BibitemOpen
  \bibfield  {author} {\bibinfo {author} {\bibfnamefont {W.~J.}\ \bibnamefont
  {Stronge}},\ }\href@noop {} {\emph {\bibinfo {title} {Impact mechanics}}}\
  (\bibinfo  {publisher} {Cambridge University Press},\ \bibinfo {address}
  {Cambridge},\ \bibinfo {year} {2004})\BibitemShut {NoStop}%
\bibitem [{\citenamefont
  {Israelachvili}(2015)}]{israelachvili2015intermolecular}%
  \BibitemOpen
  \bibfield  {author} {\bibinfo {author} {\bibfnamefont {J.~N.}\ \bibnamefont
  {Israelachvili}},\ }\href@noop {} {\emph {\bibinfo {title} {Intermolecular
  and surface forces}}}\ (\bibinfo  {publisher} {Academic Press},\ \bibinfo
  {address} {Amsterdam},\ \bibinfo {year} {2015})\BibitemShut {NoStop}%
\end{thebibliography}%

\end{document}